\documentclass[11pt,letterpaper]{article}
\usepackage{fullpage}
\usepackage{amsfonts}

\usepackage{graphicx}
\usepackage{subfigure}
\usepackage{times}
\usepackage{ifpdf}
\usepackage{url}
\usepackage{amsmath}
\renewcommand{\r}{\mathbf{r}}
\renewcommand{\P}{\mathcal{P}}
\newcommand{\V}{\mathcal{V}}
\renewcommand{\v}{\mathbf{v}}

\newcommand{\x}{\mathbf{x}}
\newcommand{\poly}{\mathrm{poly}}
\newcommand{\polylog}{\operatorname{polylog}}
\renewcommand{\a}{\mathbf{a}}
\renewcommand{\b}{\mathbf{b}}
\newcommand{\p}{p}
\newcommand{\y}{\mathbf{y}}
\newcommand{\sjs}{\ensuremath{F_2}}
\newcommand{\distinct}{\ensuremath{F_0}}
\newcommand{\matvect}{{\sc MVMult}}
\usepackage{color}
\usepackage{balance}
\newcommand{\edit}[1]{{#1}}
\newcommand{\graham}[1]{{#1}}
\newcommand{\pmw}{{\sc PMwW}}
\newcommand{\para}[1]{\medskip \noindent {\bf #1}}
\newtheorem{theorem}{Theorem}[section]
\newtheorem{lemma}[theorem]{Lemma}
\newtheorem{corollary}[theorem]{Corollary}

\newtheorem{definition}[theorem]{Definition}

\newcommand{\etal}{\textit{et al.\ }}
\newcommand{\eat}[1]{}
\newcommand{\jedit}[1]{{#1}}

\title{Practical Verified Computation with Streaming Interactive Proofs}

\author{Graham Cormode\thanks{AT\&T Labs---Research,
graham@research.att.com} 
 \and 
Michael Mitzenmacher\thanks{Harvard University, 
School of Engineering and Applied Sciences, michaelm@eecs.harvard.edu. This work was supported by NSF grants CCF-0915922,
 CNS-0721491, and IIS-0964473, and in part by grants from Yahoo! Research, Google, and
 Cisco, Inc.}
\and Justin Thaler\thanks{Harvard University, 
School of Engineering and Applied Sciences,
jthaler@seas.harvard.edu. Supported by the Department of Defense (DoD) through the National Defense Science \& Engineering Graduate Fellowship (NDSEG) Program, and in part by NSF grants CCF-0915922 and
 CNS-0721491.}}
\date{}

\renewcommand{\paragraph}[1]{\vspace{2mm}\noindent{\bf #1.}}

\begin{document}
\maketitle

\begin{abstract}
When delegating computation to a service provider, as in the cloud computing
paradigm, we
seek some reassurance that the output is correct and complete. 
Yet recomputing the output as a check is inefficient and expensive, and
it may not even be feasible to store all the data locally. 
We are therefore interested in
what can be validated by a streaming (sublinear space) user, who
cannot store the full input, or perform the full computation herself.
Our aim in this work is to advance a recent line of work on ``proof
systems'' in which the service provider {\em proves} the correctness of
its output to a user.
The goal is to minimize the time and space costs of both parties in
generating and checking the proof. 
Only very recently have there been attempts to implement such proof
systems, and thus far these have been quite limited in functionality.  

Here, our approach is two-fold. 
First, we describe a carefully chosen instantiation of one of the most efficient
general-purpose constructions for arbitrary computations 
(streaming or otherwise), due to Goldwasser,
Kalai, and Rothblum \cite{muggles}.
This requires several new insights and enhancements to 
move the methodology from a theoretical result to a practical
possibility. Our main contribution is in achieving a prover that 
runs in time $O(S(n) \log S(n))$, where $S(n)$ is the size of an arithmetic circuit computing the function of interest;
this compares favorably to the $\poly(S(n))$ runtime for the prover promised in \cite{muggles}. Our experimental results
demonstrate that a practical general-purpose protocol for verifiable computation may be significantly closer to reality than previously
realized.

Second, we describe a set of techniques that achieve
genuine scalability for protocols
fine-tuned for specific important problems in streaming and database
processing. Focusing in particular on \emph{non-interactive} protocols for problems ranging from matrix-vector multiplication
to bipartite perfect matching,
we build on prior work \cite{annotations, graphstream} to achieve a
prover that runs in nearly linear-time, 
while obtaining optimal tradeoffs between communication cost and the user's working memory. 
Existing techniques required (substantially)
superlinear time for the prover. Finally, we develop improved \emph{interactive} protocols for specific problems based on a linearization 
technique originally due to Shen \cite{shen92}. 
We argue that
even if general-purpose methods improve,
fine-tuned protocols will remain valuable in real-world settings for
key problems, and hence special attention to specific problems is warranted.
\end{abstract}

\section{Introduction}

One obvious impediment to larger-scale adoption of cloud computing
solutions is the matter of trust.  In this paper, we are specifically
concerned with trust regarding the \emph{integrity} of outsourced
computation.  If we store a large data set with a service provider,
and ask them to perform a computation on that data set, how can the
provider convince us the computation was performed correctly?
Even assuming a non-malicious  service provider,
errors due to faulty algorithm implementation, disk failures,
or memory read errors are not uncommon, especially when operating on massive 
data.  

A natural approach, which has received significant attention
\graham{particularly within the theory community}, is to require the service
provider to provide a {\em proof} along with the answer to the query.  
Adopting the terminology of proof systems \cite{arorabarak}, 
we treat the user as a verifier $\V$, who wants to
solve a problem with the help of the service provider who acts as a
prover $\P$.  
After $\P$ returns the answer, the two parties conduct a
conversation following an established protocol that satisfies the following property:
an honest prover will always convince the verifier to
accept its results, while any dishonest or mistaken prover will almost certainly
be rejected by the verifier. 
This model has led to many interesting theoretical
techniques in the extensive literature on \emph{interactive proofs}.
However, the bulk of the foundational work in this area assumed
that the verifier can afford to spend polynomial time and resources in
verifying a prover's claim to have solved a hard problem (e.g. an NP-complete problem).  
In our setting, this is too much: rather, the prover should be
efficient, ideally with effort close to linear in the input size, and the verifier should be lightweight, with effort that is
{\em sublinear} in the size of the data.

To this end, we additionally focus on results where the verifier
operates in a streaming model, taking a single pass over the
input and using a small amount of space.
This naturally fits the cloud setting, as the verifier can perform
this streaming pass while uploading the data to the cloud.
For example, consider a retailer who forwards each transaction
incrementally as it occurs. 
We model the data as too large for the user to even store in
memory, hence the need to use the cloud to store the data as it is
collected.  
Later, the user may ask the cloud to perform some
computation on the data.  The cloud then acts as a prover, sending
both an answer and a proof of integrity to the user, keeping in mind the user's space
restrictions.  

We believe that such mechanisms are vital to expand the commercial
viability of cloud computing services by allowing a trust-but-verify
relationship between the user and the service provider. Indeed, even if every
computation is not explicitly checked, the mere ability to check the
computation could stimulate users to adopt cloud computing
solutions.  Hence, in this paper, we focus on the issue of the
practicality of streaming verification protocols.  

There are many relevant costs for such protocols. In the streaming
setting, the main concern is the space used by the verifier and
the amount of communication between $\P$ and $\V$.  Other important
costs include the space and time cost to the prover, the runtime
of the verifier, and the total number of messages exchanged between
the two parties. If any one of these costs is too high, the protocol
may not be useful in real-world outsourcing scenarios.

In this work, we take a two-pronged approach.  Ideally, we would like to have a
general-purpose methodology that allows us to construct an efficient
protocol for an arbitrary computation.  
We therefore examine the costs of one of
the most efficient general-purpose protocols known in the literature on
interactive proofs,  
due to Goldwasser, Kalai, and Rothblum \cite{muggles}.  
We describe \graham{an efficient instantiation of this protocol, in which the prover is
significantly faster than in prior work,}
and present several modifications which we needed to
make our implementation scalable. 
We believe our 
\graham{success in implementing this protocol demonstrates}
that a fully practical method for reliable delegation of arbitrary
computation is much closer to reality than previously realized.

Although encouraging, our general-purpose implementation is not yet practical for everyday 
use.  Hence, our
second line of attack is to improve upon the general construction via specialized protocols for a large subset of important
problems. Here, we describe two techniques in particular that yield significantly more scalable
protocols than previously known. First, we show how to use certain Fast Fourier Transforms to obtain highly scalable 
\emph{non-interactive} protocols that are suitable for practice today; these protocols require just one message from $\P$ to $\V$,
and no communication in the reverse direction.
Second, we describe how to use a
`linearization' method applied to polynomials to obtain improved \emph{interactive} protocols for certain problems. 
All of our work is backed by empirical
evaluation based on our implementations.

Depending on the technique and the problem in question, we see
empirical results
that vary in speed by five orders of magnitude in terms of the cost to
the prover.
Hence, we argue
that even if general-purpose methods improve, fine-tuned protocols for
key problems  will remain valuable in real-world settings, especially
as these protocols can be used as primitives in more general
constructions. 
Therefore, special attention to specific problems is warranted.
%
%
The other costs of providing proofs are acceptably low. 
For many problems our
methods require at most a few megabytes of space and communication
even when the input consists of terabytes of data, and some use much
 less; moreover, the time costs of
 $\P$ and $\V$ scale linearly or almost linearly with the size of the input. 
Most of our protocols require a polylogarithmic number of messages 
between $\P$ and $\V$, but a few are non-interactive,
and send just one message. 

To summarize, we view the contributions of this paper as:
\begin{list}{\labelitemi}{\leftmargin=1em}
\item 
A carefully engineered general-purpose implementation of the
  circuit checking construction of \cite{muggles}, along with 
some extensions to this protocol. 
We believe our results show that a practical delegation protocol for arbitrary
computations is significantly closer to reality than previously realized.

\item The development of powerful and broadly applicable methods for
  obtaining practical 
specialized protocols for large classes of problems. We demonstrate
empirically that these techniques easily 
scale to streams with billions of updates.
\end{list}

\subsection{Previous Work}
\label{sec:prevwork}
The concept of an interactive proof was introduced in a burst of
activity around twenty years ago \cite{babai85,gmr85,lund92,ippspace,shen92}. 
This culminated in a celebrated result of Shamir \cite{ippspace}, which showed that the set of problems with
efficient interactive
proofs is exactly the set of problems that can be computed in polynomial
space. 
However, these results were primarily seen as theoretical statements about
computational complexity, and did not lead to implementations. 
More recently, motivated by real-world applications involving the delegation of computation, there has been
considerable interest in proving that the cloud is operating correctly.
For example,
one line of work considers methods for proving that data is being
{\em stored} without errors by an external source such as the cloud, e.g.,
\cite{proofofretrievability}.  

In our setting, we model the verifier as capable of accessing the data
only via a single, streaming pass. 
Under this constraint, there has been work in the database community
on ensuring that simple functions based on grouping and counting are
performed correctly; see
\cite{yang09:_authen} and the references therein. 
Other similar examples include work on verifying queries on a data stream with
sliding windows using Merkle trees~\cite{li07:_proof_} and verifying
continuous queries over streaming data~\cite{papadopoulos:cont}. 

Most relevant to us is work which verifies more complex and more
general functions of the input. 
The notion of a streaming verifier, who must read first the
input and then the proof under space constraints, was formalized by Chakrabarti \etal
\cite{annotations} and extended by the present authors in \cite{graphstream}. 
These works allowed the prover to send only a single message to the
verifier, with no communication in the reverse direction.
With similar motivations, Goldwasser \etal \cite{muggles} give a powerful protocol that achieves a polynomial
time prover and highly-efficient verifier for a large class of problems,
although they do not explicitly present their protocols in a streaming
setting. 
\jedit{Subsequently,} it has been noted that the information required by the verifier
can be collected with a single initial streaming pass,
and so for a large class of uniform computations, the verifier operates with
only polylogarithmic space and time. 
Finally, Cormode \etal \cite{pods} introduce the notion of
streaming interactive proofs, extending the model of
\cite{annotations} by allowing multiple rounds of interaction between
prover and verifier. They
present exponentially cheaper protocols than those possible in the single-message model
of \cite{annotations,graphstream},
for a variety of problems of central
importance in database and stream processing.  

A different line of work has used fully homomorphic encryption to ensure integrity,
privacy, and reusability in delegated computation \cite{ggp, kai-min1, kai-min2}.
The work of Chung, Kalai, Liu, and Raz \cite{kai-min2} is particularly related, as they
focus on delegation of streaming computation. 
Their results are stronger than ours, in that they achieve reusable general-purpose protocols (even if $\P$
learns whether $\V$ accepts or rejects each proof), but their soundness guarantees rely on computational assumptions,
and the substantial overhead due to the 
use of fully homomorphic encryption
means that these protocols remain far
from practical at this time.

Only very recently have there been sustained efforts to use 
techniques derived from the complexity and cryptography
worlds to actually verify computations. Bhattacharyya implements certain PCP constructions 
and indicates they may be close to practical \cite{techreport}.
In parallel to this work, Setty \etal~\cite{hotos, ndss} survey
results on probabilistically checkable proofs (PCPs),
and implement a construction originally due to Ishai \etal~\cite{ishai}. While their work represents a clear advance in the
implementation of PCPs, our approach has
several advantages over \cite{hotos, ndss}.
For example, our protocols save space and time for the verifier even when outsourcing a single computation, while \cite{hotos, ndss} saves time for the verifier only when batching 
together several dozen computations at once and amortizing the verifier's cost over the batch. Moreover, our protocols are unconditionally secure even against computationally unbounded 
adversaries, while the construction of Ishai \etal relies on cryptographic assumptions to obtain security guarantees. 
%
Another \graham{practically-motivated} approach is due to Canetti \etal \cite{riva}. Their implementation 
delegates the computation
to two {\em independent} provers, and ``plays them off'' against each other: if they
disagree on the output, the protocol identifies where their executions diverge, and favors the one which follows the program
correctly at the next step. 
This approach requires at least one of the provers to be honest for
any security guarantee to hold.

\subsection{Preliminaries}
\label{sec:prelims}
\para{Definitions.}
We first formally define a valid protocol.  
Here we closely follow previous work, such as \cite{pods} and
\cite{annotations}.

\noindent \begin{definition}
Consider a prover $\P$ and verifier $\V$ who both observe a stream
$\mathcal{A}$ and wish to compute a function  
$f(\mathcal{A})$.
After the stream is observed, $\P$ and $\V$ exchange a sequence of messages.
Denote the output of
$\V$ on input $\mathcal{A}$, given prover
$\P$ and $\V$'s random bits $\mathcal{R}$, by 
$\operatorname{out}(\V, \mathcal{A}, \mathcal{R}, \P)$.  
$\V$ can  output $\perp$ if $\V$ is not convinced that $\P$'s claim is valid.

$\P$ is a \textit{valid prover} with respect to $\V$ 
if for all streams $\mathcal{A}$,
$$\Pr_{\mathcal{R}}[\operatorname{out}(\V, \mathcal{A}, \mathcal{R},
\P) = f(\mathcal{A})] =1.$$
We call $\V$ a valid verifier for $f$ if 
there is at least one valid prover $\P$  with respect to $\V$, 
and for all provers $\P'$ and all streams $\mathcal{A}$,
$$\Pr_{\mathcal{R}}[\operatorname{out}(\V, \mathcal{A}, \mathcal{R},
\P') \not\in \{f(\mathcal{A}),\perp\}] \leq 1/3.$$ 
\end{definition}

Essentially, this definition states that a prover who follows
the protocol correctly will always convince $\V$, while if $\P$ makes
any mistakes or false claims, then this will be detected with at least
constant probability.  In fact, for our protocols, this
  `false positive' probability can easily be made arbitrarily small.


As our first concern in a streaming setting is the space
requirements of the verifier as well as the communication cost for the
protocol, we make the following definition. 
\begin{definition}
\label{def:protocol} 
We say $f$ possesses an $r$-message $(h, v)$ protocol, if there exists
a valid verifier $\V$ for $f$ such that: 

1. $\V$ has access to only $O(v)$ words of working memory. 

2. There is a valid prover $\P$ for $\V$ such that $\P$  and $\V$
exchange at most $r$ messages in total, and
the sum of the lengths of all messages is $O(h)$ words. 
\end{definition}
We refer to one-message protocols as \emph{non-interactive}. 
We say an $r$-message protocol has $\lceil r/2 \rceil$ \emph{rounds}. 

A key step in many proof systems is the evaluation of the 
{\em low-degree extension} of some data at multiple points. 
That is, the data is interpreted as implicitly defining a polynomial
function which agrees with the data over the range $1\ldots n$,
and which can also be evaluated at points outside this range as a check. 
The existence of streaming verifiers relies on the fact that such
low-degree extensions can be evaluated at any given location
incrementally as the data is presented \cite{pods}. 

\edit{
\medskip
\noindent \textbf{Input Representation.} 
\graham{
All protocols presented in this paper can handle inputs specified in a
very general data stream form.}
Each element of the stream is a tuple $(i, \delta)$, where $i \in [n]$
and $\delta$ is an  
integer (which may be negative, thereby modeling deletions). The data stream
implicitly defines a \emph{frequency vector} $a$, where $a_i$ is the sum
of all $\delta$ values associated with $i$ in the stream, and the goal
is to compute a function of $a$. 
\graham{
Notice the function of $a$ to be computed may interpret $a$ as an
object other than a vector, such as a matrix or a string. 
For example, in the \matvect\ problem described below, 
$a$ defines a matrix and a vector to be multiplied, 
and in some of the graph problems considered as
extensions in Section \ref{sec:FFT}, $a$ defines the adjacency matrix
of a graph. }

In Sections \ref{sec:FFT} and \ref{sec:muggles}, the manner in which
we describe protocols may appear to assume that the data stream has
been pre-aggregated into the frequency vector $a$ (for example, in
Section \ref{sec:muggles}, we apply the protocol of Goldwasser \etal \cite{muggles} to arithmetic circuits whose $i$'th input wire has
value $a_i$). 
It is therefore important to emphasize that in fact all
of the protocols in this paper can be executed in the input model of
the previous paragraph, where $\V$ only 
sees the raw (unaggregated) stream and not the aggregated frequency
vector $a$, and there is 
no explicit conversion between the raw stream and the aggregated
vector $a$. 
\graham{ This follows from observations in \cite{annotations, pods},
which we describe here for completeness.}

The critical observation is that in all of our protocols, the only
information $\V$ must extract from the data stream is the evaluation of a
low-degree extension of $a$ at a random point $\r$, which we denote by
$\text{LDE}_a(\r)$, and this  
value can be computed incrementally by $\V$ using $O(1)$ words of
memory as the raw stream is presented to $\V$. 
\graham{Crucially this is possible because, }
for fixed $\r$, the function $a
\mapsto \text{LDE}_a(\r)$ is \emph{linear}, and thus it is
straightforward for $\V$ to compute the contribution of each update
$(i, \delta)$ to $\text{LDE}_a(\r)$.  

More precisely, we can write
$\text{LDE}_a(\r) = \sum_{i \in [n]} a_i \chi_i(\r)$, where $\chi_i$
is a (Lagrange) polynomial that depends only on $i$. 
Thus, $\V$ can compute $\text{LDE}_a(\r)$ 
incrementally from the raw stream by initializing $\text{LDE}_a(\r) \leftarrow 0$,
and processing each update $(i, \delta)$ via:
$$\text{LDE}_a(\r) \leftarrow \text{LDE}_a(\r) + \delta \chi_i(\r).$$ 
$\V$ only needs to store $\text{LDE}_a(\r)$ and $\r$, which requires $O(1)$ words of memory. Moreover, for any $i$, $\chi_i(\r)$ can be computed in $O(\log n)$ field operations, and thus $\V$ can compute $\text{LDE}_a(\r)$ with one pass over the raw stream, using $O(1)$ words of space and $\log n$ field operations per update.}

\para{Problems.}
\label{sec:problems}
To focus our discussion and experimental study, we describe four key
problems that capture different aspects of computation: data grouping
and aggregation, linear algebra, and pattern matching.  
We will study how to build valid protocols for each of these
problems. 
Throughout, let $[n]=\{0, \dots, n-1\}$ 
denote the universe from which data elements are drawn.

\begin{trivlist}
\item
\sjs:
Given a stream of $m$ elements from $[n]$, compute
$\sum_{i\in [n]}  a_i^2$ where $a_i$ is the number of occurrences of $i$ in
the stream.  
This is also known as the {\em second frequency moment}, a special
case of the $k$th frequency moment $F_k = \sum_{i\in [n]} a_i^k$. 
\item \distinct:
Given a stream of $m$ elements from $[n]$, compute
the number of {\em distinct} elements, i.e. 
the number of $i$ with $a_i > 0$, where again $a_i$ is the number of
occurrences of $i$ in the stream. 
\item
\matvect:
Given a stream defining an $n \times n$ integer matrix $A$, 
and vectors $\mathbf{x}, \mathbf{b} \in \mathbb{Z}^n$, determine
whether $A\mathbf{x} = \mathbf{b}$.
More generally, we are interested in the case where $\P$ provides 
a vector $\mathbf{b}$ which is claimed to be $A\mathbf{x}$. 
This is easily handled by our protocols, since $\V$ can treat 
the provided $\mathbf{b}$ as part of the input, even though it may arrive
after the rest of the input. 
\eat{
Instead, we show how to obtain communication linear in $n$, by 
 describing a circuit for {\em verifying} that $A\mathbf{x}=\mathbf{b}$,
where $\mathbf{b}$ is a vector that can be sent to $\V$ by $\P$ after
$\V$ has observed the input. 
That is, the input to the circuit consists of the matrix $A$, the
vector $\mathbf{x}$ {\em and} the vector $\mathbf{b}$ which is claimed
to be $A\mathbf{x}$. 
Note that, because of the flexible way in which $\V$ computes the
check on the input, it is straightforward for $A$ and $\mathbf{x}$ to
be derived from a stream while $\mathbf{b}$ is given later by $\P$. }

\item
\pmw:
Given a stream representing text $T = (t_0, \dots , t_{n-1}) \in
[n]^n$ and pattern $P= (p_0, \dots , p_{q-1})\in[n]^q$, 
the pattern $P$ is said to
occur at location $i$ in $t$ if, for every position $j$ in $P$, either $p_j = t_{i+j}$
or at least one of $p_j$ and $t_{i+j}$ is the wildcard symbol $*$. The pattern-matching with wildcards problem
is to determine the number of locations at which $P$ occurs in $T$. 
\end{trivlist}

For simplicity, we will assume the stream length $n$ and the universe size $m$
are on the same order of magnitude i.e. $m = \Theta(n)$.

All four problems require linear space in the
streaming model to solve exactly (although there are
space-efficient approximation algorithms for the first three
\cite{Muthukrishnan:03}).  

\medskip
\noindent \textbf{Non-interactive versus Multi-round Protocols.}
\eat{
There are two definition allows protocols where the prover sends only a single
message to the verifier, without any other communication between them
(the non-interactive case), or where both players have a sustained
conversation, possibly spanning hundreds of rounds or more (the
multiround).}
Protocols for reliable delegation fall into two classes:
non-interactive, in which a single message is sent from prover to verifier
and no communication occurs in the reverse direction; and multi-round, 
where the two parties have a sustained
conversation, possibly spanning hundreds of rounds or more.
There are merits and drawbacks to each.

\medskip
\noindent \emph{--- Non-interactive Advantages}: The non-interactive model has the 
desirable property that the prover can
compute the proof and send it to the verifier (in an email, or posted
on a website) for $\V$ to retrieve and validate at her leisure. 
In contrast, the multi-round case requires $\P$ and $\V$ to interact
online. 
Due to round-trip delays, the time cost of multi-round protocols can
become high; moreover, $\P$ may have to do substantial computation after each
message.
This can involve maintaining state between messages, and performing
many passes over the data. 
\graham{A less obvious advantage is that non-interactive protocols can be
\jedit{repeated} for different instances (e.g. searching for different
patterns in \pmw) without requiring $\V$ to use fresh randomness. This allows the verifier to amortize
much of its time cost over many queries, potentially achieving sublinear time cost per query.
The reason this is possible is that in the course of a non-interactive protocol, $\P$ learns nothing
about $\V$'s private randomness 
(assuming $\P$ does not learn whether $\V$ accepts or rejects the proof) and so we
can use a union bound to bound the probability of error over multiple
instances. 
In contrast, in the multi-round case, $\V$ must divulge most of its
private random bits to $\P$ over the course of the protocol.}

\medskip
\noindent \emph{--- Multi-round Advantages}: The overall cost in a multi-round protocol can be lower, as
most non-interactive protocols require $\V$ to use substantial space and read a large proof.
Indeed, prior work \cite{annotations, graphstream} has shown that space or communication must be $\Omega(\sqrt{n})$ for most non-interactive protocols
\cite{annotations}. Nonetheless, even for terabyte streams of data, these costs typically
translate to only a few megabytes of space and communication, which is tolerable in many applications.  
Of more concern is that the time cost to the prover in known non-interactive protocols is
typically
much higher than in the interactive case, \edit{though this gap is not known to be inherent}.
We make substantial progress in closing this gap in prover runtime 
in Section~\ref{sec:FFT}, but this still leaves an order of magnitude
difference in practice (Section~\ref{sec:experiments}).

\subsection{Outline and Contributions}
We consider non-interactive protocols first, and interactive protocols second. 
 To begin, we describe in Section \ref{sec:FFT} how to use Fast Fourier
Transform methods to engineer $\P$'s runtime in the \sjs\ protocol of \cite{annotations} down from $O(n^{3/2})$ to
nearly-linear time.  The \sjs\ protocol is a key
target, because (as we describe) several protocols build directly upon it.
We show in Section \ref{sec:experiments} that this results in a
speedup of hundreds of thousands of updates per second, bringing this
protocol, as well as those that build upon it, from theory to practice. 

Turning to interactive protocols, in Section \ref{sec:pattern} we
describe 
\graham{an efficient instantiation of}
the general-purpose construction of \cite{muggles}. 
Here, we also describe efficient protocols for specific problems of high interest including 
\distinct\ and \pmw\ based on an application of our implementation to
carefully chosen circuits. 
The latter protocol enables verifiable
searching (even with wildcards) in the cloud, and complements 
work on searching in encrypted data within the cloud (e.g. \cite{boneh}).  
Our final contribution in this section is to demonstrate that the use of more general
arithmetic gates to enhance the basic protocol of \cite{muggles}
allows us to significantly decrease prover time, communication cost,
and message cost of these two protocols in practice.  

In Section \ref{sec:distinct} we provide alternative interactive protocols for important specific problems based on a technique known as linearization;
we demonstrate in Section \ref{sec:experiments} that linearization yields a protocol for \distinct\ in which $\P$ runs nearly two orders of magnitude 
faster than in all other known protocols for this problem. 
\graham{Finally, we describe our observations on implementing these
  different methods, including our carefully engineered implementation of the
powerful general-purpose construction of \cite{muggles}.}

\newcommand{\Fp}{\ensuremath{ \mathbb{F}_p}}
\newcommand{\Finf}{\ensuremath{F_\infty}}

\section{Fast Non-interactive Proofs via Fast Fourier Transforms}\label{sec:FFT}
In this section, we describe how to drastically speed up
$\P$'s computation for a large class of \emph{specialized},
non-interactive protocols. 
In non-interactive proofs, $\P$ often needs to evaluate a low-degree
extension at a large number of locations, which can be the
bottleneck. 
Here, we show how to reduce the cost of this step to near
linear, via Fast Fourier Transform (FFT) methods.  

For concreteness, we describe the approach in the context of a
non-interactive protocol for \sjs\ given in \cite{annotations}.  
Initial experiments on this protocol identified 
the prover's runtime
as the principal bottleneck in the protocol \cite{pods}. 
In this implementation,
$\P$ required $\Theta(n^{3/2})$ time, and consequently the
implementation fails to scale for $n > 10^7$.
Here, we show that FFT techniques can dramatically speed up
the prover, leading to a protocol that  
easily scales to streams consisting of billions of items.

We point out that \sjs\ is a problem of significant interest, 
beyond being a canonical streaming problem. 
Many existing protocols in the
non-interactive model are built on top of \sjs\
protocols, including finding the inner product and Hamming distance
between two vectors \cite{annotations}, 
the \matvect\ problem, solving a large class of linear programs, and graph problems such as testing connectivity
and identifying bipartite perfect matchings
\cite{annotationsjournal,graphstream}. These protocols are particularly important because
they all achieve
provably optimal tradeoffs between space and communication costs \cite{annotations}.
Thus, by developing a scalable, practical protocol for 
\sjs, we also achieve big improvements in protocols for a 
host of important (and seemingly unrelated) problems.

\para{Non-interactive \sjs\ and \matvect\ Protocols.}
\label{sec:oldnip}
We first outline the protocol from \cite[Theorem
  4]{annotations} for \sjs\ on an $n$ dimensional vector. This construction yields an $(n^{\alpha}, n^{1-\alpha})$ protocol
for \emph{any} $0 \leq \alpha \leq 1$, i.e. it allows a tradeoff
between the amount of communication and space used by $\V$; for brevity
we describe the protocol when $\alpha=1/2$.

Assume for simplicity that $n$ is a perfect square. 
We treat the $n$ dimensional vector as a 
$\sqrt{n} \times \sqrt{n}$ array $a$.
This implies a two-variate polynomial $f$ 
over a suitably large finite field $\Fp$, 
such that 
\[ \forall (x, y) \in [\sqrt{n}] \times [\sqrt{n}] : 
f(x,y) = a_{x,y} .\] 
To compute \sjs, we wish to compute 
\[ \sum_{x \in [\sqrt{n}], y \in [\sqrt{n}]} a^2_{x,y} = \sum_{x \in [\sqrt{n}], y \in
  [\sqrt{n}]} f^2(x,y).\]
  
The low-degree extension 
$f$ can also be evaluated at locations outside 
$[\sqrt{n}] \times [\sqrt{n}]$.
In the protocol, 
the verifier $\V$ picks a random position $r \in \Fp$, and evaluates 
$f(r,y)$ for every $y \in [\sqrt{n}]$ (\cite{annotations} shows how
$\V$ can compute any $f(r,y)$ incrementally in constant space).
The proof given by $\P$ is in the form of  
a degree $2(\sqrt{n}-1)$ polynomial $s(X)$ which is
claimed to be  $\sum_{y\in [\sqrt{n}]} f(X,y)^2$.
$\V$ uses the values of $f(r,y)$ to check that 
$s(r) = \sum_{y\in [\sqrt{n}]}f(r,y)^2$, and if so
accepts $\sum_{x \in [\sqrt{n}]} s(x)$ as the correct answer. Clearly $\V$'s check will
pass if $s$
is as claimed.
The proof of validity follows from the Schwartz-Zippel lemma:
if $s(X) \neq \sum_{y\in [\sqrt{n}]} f(X,y)^2$ as claimed by $\P$, then
\[ \Pr\big[s(r) =  \sum_{y\in [\sqrt{n}]} f(r,y)^2\big] \leq 
\frac{\operatorname{degree}(s)}{|\Fp|} = 
\frac{2(\sqrt{n}-1)}{p}\]
 where $p$ is the size of the finite field $\Fp$. Thus, if $\P$ deviates at 
 all from the prescribed protocol, the verifier's check will fail with high probability.

A non-interactive protocol for \matvect\ uses similar ideas. 
Each entry in the output is the result of an inner product between two
vectors: a row of matrix $A$ and vector $\mathbf{x}$. 
Each of the $n$ entries in the output can be checked independently with a variation of the
above protocol, where the squared values are replaced by products of
vector entries; this naive approach yields an $(n^{3/2}, n^{3/2})$ protocol for \matvect.
\cite{graphstream} observes that, because $\mathbf{x}$ is held
constant throughout all $n$ inner product computations, 
$\V$'s space requirements can be reduced by having $\V$ keep track of 
hashed information, rather than full vectors. 
The messages from $\P$ do not change, however, and computing
low-degree extensions of the input data is the chief scalability 
bottleneck.  
This construction yields a 1-message $(n^{1+\alpha}, n^{1-\alpha})$
protocol (as in Definition~\ref{def:protocol})
for any $0 \leq \alpha \leq 1$, and this can be shown to be optimal.

\subsection{Breaking the bottleneck}  
Since $s(X)$ has degree at most $2\sqrt{n}-1$
 it is uniquely specified by its values at any $2\sqrt{n}$ locations. 
We show how $\P$ can quickly evaluate all values in the set 
\[S:=\{(x, s(x)): x \in [2\sqrt{n}]\}.\]
Since $s(X) =  \sum_{y\in [\sqrt{n}]} f(X,y)^2$, given
all values in  set 
\[T:=\{(x, y, f(x, y)): x \in [2\sqrt{n}], y \in
[\sqrt{n}]\},\] 
all values in $S$ can be computed in time linear in $n$. 
The implementation
of \cite{pods} calculated each value in $T$ independently, requiring
$\Theta(n^{3/2})$ time overall. 
We show how FFT techniques allow us to calculate $T$ much faster.

The task of computing $T$ boils down to multi-point evaluation of the
polynomial $f$.  
It is known how to perform fast multi-point
evaluation of univariate degree $t$ polynomials 
in time $O(t \log t)$,
and bivariate polynomials in subquadratic time,
if the polynomial is specified by its coefficients
\cite{esa}. 
However, there is substantial overhead in converting $f$ to a
coefficient representation.
It is more efficient for us to directly work with and
exchange polynomials in an implicit representation, by specifying
their values at sufficiently many points. 

\para{Representing as a convolution.}
We are given the values of $f$ at all points located on
the $[\sqrt{n}] \times [\sqrt{n}]$ ``grid''. 
We leverage this fact to compute
$T$ efficiently in nearly linear time by a direct application of the
Fast Fourier Transform. 
For $(x, y) \in [\sqrt{n}] \times [\sqrt{n}]$,
$f(x, y)$ is just $a_{x, y}$, which $\P$ can store explicitly while
processing the stream. 
It remains to calculate 
$(x, y, f(x, y))$ for $\sqrt{n} \leq x < 2 \sqrt{n}$.
For fixed $y \in [\sqrt{n}]$, we may write $f(X, y)$ explicitly as
$$ f(X, y) = \sum_{i \in [\sqrt{n}]} a_{i, y} \chi_i(X),$$ where
$\chi_i$ is the Lagrange polynomial\footnote{
That is, the unique polynomial of degree $\sqrt{n}$ such that 
$\chi_i(i) = 1$, while for $j\neq i \in [\sqrt{n}]$, 
$\chi_i(j)=0$.
Here, the inverse is the multiplicative inverse within the field.}
\[ 
  \chi_{i}(j) = \prod_{x \in [\sqrt{n}]\setminus\{i\}} (j-i)(x-i)^{-1}\]

If $j \not\in [\sqrt{n}]$, then we may write 
\begin{align}
f(j, y) = & {\sum_{i \in [\sqrt{n}]}  h(j) b_y(i) g(j-i) }
\label{eq:conv}
\\
\nonumber
\text{where}\quad  
  b_y(i) & = { a_{i,y} \prod_{x \in [\sqrt{n}]\setminus\{i\}} (x-i)^{-1},} \\
\nonumber
h(j) & = {\prod_{k=(j +1 - \sqrt{n})}^{j} k}, \\
\nonumber
\text{and}\quad   g(j-i) &=  (j-i)^{-1}.
\end{align}

As a result $f(j,y)$ can be computed as a circular convolution of $b_y$ and
$g$, scaled by $h(j)$. 
That is, for a fixed $y$, all values in the set
$T_y := \{(x, y, f(x, y)): x \in [2\sqrt{n}]\}$ 
can be found by computing the
convolution in Equation \ref{eq:conv}, then scaling each entry by the
appropriate value of $h(j)$.


\para{Computing the Convolution.}
We represent  
$b_y$ and $g$ by vectors of length $2 \sqrt{n}$ over a suitable field, 
and take the Discrete Fourier Transform (DFT) of each.
The convolution is the inverse transform of the inner product of the
two transforms \cite[Chapter 5]{KandT}. 
%
There is some freedom to choose the field over which to perform the
transform. 
We can compute
the DFT of $f_y$ and $g$ over the complex field $\mathbb{C}$
using $O(\sqrt{n} \log n)$ arithmetic operations via standard
techniques such as the Cooley-Tukey algorithm \cite{tukey}, 
and simply reduce the final result modulo $p$, rounded to the nearest integer. 
Logarithmically many bits of precision 
past the decimal point suffice to obtain a
sufficiently accurate result.  
Since we compute $O(\sqrt{n})$ such convolutions, we obtain
the following result: 
\begin{theorem} \label{thm:F2}
The honest prover in the \sjs\ protocol of \cite[Theorem
    4]{annotations} requires $O(n \log n)$ arithmetic operations on
  numbers of bit-complexity $O(\log n + \log p)$. 
\end{theorem} 

In practice, however, working over $\mathbb{C}$ can be slow, and requires us
to deal with precision issues. 
Since the original data resides in some finite field \Fp, and
can be represented as fixed-precision integers, it is preferable to 
also compute the DFT over the same field. 
Here, we exploit the fact that in designing our protocol, we can
choose to work over \emph{any} sufficiently large finite field \Fp.

There are two issues to address: 
we need that there {\em exists} a DFT for sequences of length
$2\sqrt{n}$ (or thereabouts) in \Fp, and further that this 
DFT has a corresponding ({\em fast}) Fourier Transform algorithm. 
We can resolve both issues with the \emph{Prime Factor Algorithm} (PFA) 
for the DFT in \Fp~\cite{PFA}. 
The ``textbook'' Cooley-Turkey FFT algorithm operates on sequences whose length
is a power of two.  
Instead, the PFA works on sequences of 
length $N=N_1 \times N_2 \times \ldots \times N_k$, 
where the $N_i$'s are pairwise coprime.
The time cost of the transform is $O((\sum_i N_i) N)$. 
The algorithm is typically applied over the complex numbers, but 
 also applies over $\mathbb{F}_p$:  
 it works by breaking the large DFT up into a sequence of smaller
 DFTs, each of size $N_i$ for some $i$.
These base DFTs for sequences of length $N_i$ exist for \Fp
whenever there exists a primitive $N_i$'th root of unity in \Fp.
This is the case whenever $N_i$ is a divisor of $p-1$. 
So we are in good shape so long as $p-1$ has many distinct prime
factors. 

Here, we use our freedom to fix $p$, and choose
$p=2^{61}-1$.\footnote{Arithmetic in this field can also be 
  done quickly, see Section~\ref{sec:implementation}.}
Notice that  
$$2^{61}-2 = 2 \times 3^2 \times 5^2 \times 7 \times 13 \times 31
\times 41 \times 61 \times 151 \times 331 \times 1321,$$
and so there are many such divisors $N_i$ to choose from
when working over $\mathbb{F}_p$. 
If $2 \sqrt{n}$ is not equal to a factor of $p-1$, we can simply pad
the vectors $f_y$ and $g$ 
such that their lengths are factors of $2^{61}-2$. 
Since $2^{61}-2$ has many \emph{small} factors, we never have to use
too much padding: 
we calculated that we never need to pad any sequence of length
$100 \leq N \leq 10^9$ (good for $n$ up to $10^{18}$)
by more than $16\%$ of its length.
This is better than the Cooley-Tukey method, where padding can
double the length of the sequence. 

As an example, we can work with the length 
$N = 2\times 5 \times 7 \times 9 \times 11 \times 13 = 90090$, 
 sufficient for inputs of size $n = (N/2)^2$, which is over $10^9$. 
The cost scales as $(2+5+7+9+11+13)N = 47N$. 
Therefore, the PFA approach 
offers a substantial improvement over naive convolution in
$\mathbb{F}_p$, which takes time $\Theta(N^2)$. 

\para{Parallelization.}
This protocol is highly amenable to parallelization. 
Observe that $\P$ 
performs $O(\sqrt{n})$ independent convolutions of each of length
$O(\sqrt{n})$  (one for each column $y$ of the matrix $a_{x, y}$), 
followed by computing $\sum_y a_{x, y}^2$ for each row $x$ of the result. 
The convolutions can be done in parallel, and once complete, the
sum of squares of each row can also be parallelized.
This protocol also possesses a simple two-round MapReduce
protocol. 
In the first round, we assign each column $y$ of the
matrix $a_{x,y}$ a unique key, and have each reducer perform the
convolution for the corresponding column. In the second round, we
assign each row $x$ a unique key, and have each reducer compute
$\sum_y a_{x,y}^2$ for its row $x$. 

\edit{\subsection{Implications}
As we experimentally demonstrate in Section \ref{sec:experiments}, the
results of this section make practical the fundamental building block
for the majority of known non-interactive protocols. 
Indeed, by combining Theorem \ref{thm:F2} with protocols 
from \cite{annotations,graphstream}, 
we obtain the following immediate corollaries. 
For all graph problems considered, $n$ is the number of nodes in the
graph, and $m$ is the number of edges.  

\begin{corollary}
\label{cor:F2}
\begin{enumerate} 
\item (Extending \cite[Theorem 4.3]{annotations}) For any $h \cdot v \geq n$, there is an $(h, v)$ protocol for computing the inner product and Hamming distance of two $n$-dimensional vectors, where $\V$ runs in time $O(n)$ and $\P$ runs in time $O(n \log n)$. The previous best runtime known for $\P$ was $O(h^2v)$.
\item (Extending \cite[Theorem 4]{graphstream}) For any $h \cdot v
  \geq n$, there is an $(mh, v)$ protocol for $m \times n$ integer
  matrix-vector multiplication (\matvect), where $\V$ runs in time $O(mn)$ and $\P$ runs in time $O(mn \log n)$. The best runtime known for $\P$ previously was $O(mh^2v)$. 
\item (Extending \cite[Corollary 3]{graphstream}) For any $h \cdot v \geq n$, there is an $O(nh, v)$ protocol for 
solving a linear program over $n$ variables with $n$ (integer) constraints and subdeterminants of polynomial magnitude, where $\V$ runs in time $O(n^2)$ and $\P$ runs in time
$O(t(n) + n^2 \log n)$, where $t(n)$ is the time required to solve the linear program and its dual.  The best runtime known for $\P$ previously was $O(t(n) + nh^2v)$. 
\item (Extending \cite[Theorem 5.4]{annotations}) For any $h \cdot v \geq n^3$, there is an $(h, v)$ protocol for counting the number of triangles 
in a graph, where $\V$ runs in time $O(mn)$ and $\P$ runs in time $O(n^3 \log n)$. The best runtime known for $\P$ previously was $O(h^2 v)$. 
\item (Extending \cite[Theorem 6.6]{annotationsjournal}) For any $h \cdot v \geq n^2$, $h \geq n$, there is an $(h, v)$ protocol for graph connectivity, where $\P$ runs in time $O(n^2 \log n)$ and $\V$ runs in time $O(m)$. The best runtime known for $\P$ previously was $O(nh^2v)$. 
\item (Extending \cite[Theorem 6.5]{annotationsjournal}) For any $h \cdot v \geq n^2$, $h \geq n$, there is an $(h, v)$
  protocol for bipartite perfect matching, where $\V$ runs in time
  $O(m)$ and $\P$ runs in time $O(t(n) + n^2 \log n)$, 
where $t(n)$ is the time required to find a perfect matching if one
exists, \graham{or to find a counter-example (via Hall's Theorem)
  otherwise. }
The best runtime known for $\P$ previously was $O(t(n) + h^2v)$. 
\end{enumerate}
\end{corollary}

\graham{
In the common case where we choose $h = v$, this represents a
polynomial-speed up in $\P$'s runtime.
For example, for the \matvect\ problem, the prover's cost is reduced from
$O(mn^{3/2})$ in prior work to $O(mn \log n)$. }

In most cases of Corollary \ref{cor:F2}, $\V$ runs in linear time, and
$\P$ runs in nearly linear time for dense inputs, plus the time $t(n)$
required to solve the problem in the first place, which may be
superlinear. 
Thus, $\P$ pays at most a logarithmic factor overhead in solving the
problem ``verifiably'', compared to solving the problem in a
non-verifiable manner.  }
\eat{
The \matvect\ protocol is even easier to
parallelize, as this protocol has the prover run an independent
inner-product protocol for each row of the matrix. 
In the unlikely event there are more processors available than rows of
the matrix, the protocol can be further parallelized by applying the
technique of the previous paragraph to each row of the matrix.  
}
 
\eat{
As mentioned above, these techniques extend straightforwardly to yield a more efficient protocol for {\sc Matrix-vector multiplication} and a large class of linear programs \cite{graphstream},
and efficient protocols for {\sc Graph Connectivity} and {\sc Bipartite Perfect Matching} that use {\sc Matrix-vector multiplication} as a subroutine \cite{annotationsjournal}. They also additionally yield improved protocols for {\sc Inner Product} and {\sc Hamming Distance} \cite{annotations}. 
}

{\fussy
\section{A General Approach: Multi-round Protocols Via Circuit Checking}\label{sec:muggles}}
\label{sec:pattern}
In this section, we study \emph{interactive} protocols, 
and describe how to \graham{efficiently instantiate the
 powerful framework} due to Goldwasser, Kalai, and Rothblum for verifying
arbitrary computations\footnote{We are indebted to these authors for
  sharing their working draft of the full version of \cite{muggles},
  which provides much greater detail than is possible in the conference
  presentation.}.

A standard approach to verified computation developed in the theoretical
literature is to verify properties of circuits
that compute the desired function \cite{ggp,muggles,hotos}. 
One of the most promising of these is due to Goldwasser {\em et al.}, which
proves the following result: 

\begin{theorem}\label{thm:muggles} \cite{muggles} Let $f$ be a function over an arbitrary field $\mathbb{F}$ that can be computed
by a family of $O(\log S(n))$-space uniform arithmetic circuits (over
$\mathbb{F}$) of fan-in 2, 
size $S(n)$, and depth $d(n)$. Then, assuming unit cost for
transmitting or storing a value in $\mathbb{F}$,  
$f$ possesses a $(\log S(n), d(n)\log S(n))$-protocol requiring $O(d(n)\log S(n))$ rounds.
$\V$ runs in time $\left(n + d(n)\right) \polylog\left(S(n)\right)$ and $\P$ runs in time $\poly(S(n))$.
\end{theorem}

Here, an arithmetic circuit over a field $\mathbb{F}$ is analogous to a boolean circuit, except 
that the inputs are elements of $\mathbb{F}$ rather than boolean values, and the gates of the circuit
compute addition and multiplication over $\mathbb{F}$.
We address how to realize the protocol of Theorem~\ref{thm:muggles}
efficiently. \edit{Specifically, we show three technical results. The first two results, Theorems \ref{OPT1} and \ref{thm:online}, state that
for \emph{any} log-space uniform circuit, the honest prover in the 
protocol of Theorem \ref{thm:muggles} 
can be made to run in time nearly linear in the size of the circuit, with a streaming verifier
who uses only $O(\log S(n))$ words of memory. Thus, these results guarantee a highly efficient prover and a space-efficient verifier. 
In streaming contexts, 
where $\V$ is more space-constrained than 
time-constrained, this may be acceptable. Moreover, Theorem \ref{thm:online} states that $\V$ can perform the time-consuming part of its computation in a data-independent non-interactive preprocessing phase, which can occur offline before the stream is observed.

Our third result, Theorem \ref{OPT2} makes a slightly stronger assumption but yields a stronger result: it states that 
under very mild conditions on the
circuit, we can achieve a prover who runs in time nearly linear in the size of the circuit,
and a verifier who is \emph{both} space- and time-efficient.


Before stating our theorems, we sketch the main techniques needed to
achieve the efficient implementation, with full details in Appendix
\ref{fullapp:muggles}. 
We also direct the interested reader to the source code of our
implementations \cite{code}. 
The remainder of this section
is intended to be reasonably accessible to readers who are familiar
with the sum-check protocol \cite{ippspace, lund92}, 
but not necessarily with the protocol of \cite{muggles}.
}

\subsection{Engineering an Efficient Prover} 
In the protocol of \cite{muggles}, $\V$ and $\P$ first agree on a
depth $d$ circuit $C$ of gates with fan-in 2 
that computes the function of interest; $C$ is assumed
to be in layered form \edit{(this assumption blows up the size of the circuit by at most a factor
of $d$, and we argue that it is unrestrictive in practice, as the natural circuits for all four of our motivating problems are layered, 
as well as for a variety of other problems described in Appendix
\ref{fullapp:muggles}).}  
$\P$ begins by claiming a value for the output
gate of the circuit. The protocol then proceeds iteratively from the
output layer of $C$ to the input layer, with one iteration for each
layer. 
For presentation purposes, assume that all layers of the circuit have $n$
gates, and let $v=\log n$.

\edit{At a high level, in iteration $1$, $\V$ reduces verifying the claimed
value of the output gate to computing the value of a certain $3v$-variate
polynomial $f_1$ at a random point $r^{(1)} \in \mathbb{F}^{3v}$.
The iterations then proceed inductively over each layer of gates: 
in iteration $i>1$, $\V$ reduces computing
$f_{i-1}(r^{(i-1)})$ for a certain $3v$-variate polynomial $f_{i-1}$ to computing 
$f_{i}(r^{(i)})$ for 
a random point $r^{(i)}\in \mathbb{F}_p^{3v}$.
Finally, in iteration $d$, $\V$ must compute
$f_d(r^{(d)})$.
This happens to be a function of the input alone (specifically, it is an
evaluation of a low-degree extension of the input), 
and $\V$ can compute this value in a streaming fashion, without assistance,
even if only given access to the raw (unaggregated) data stream, as described
in Section \ref{sec:prelims}.
If the values agree, then $\V$ is convinced of the correctness of the
output.

We abstract the notion of a ``wiring predicate'',  
which encodes which pairs of wires from
layer $i-1$ are connected to a given gate at layer $i$.
Each iteration $i$ consists of an application of the standard
\emph{sum-check protocol} \cite{lund92, ippspace}
to a $3v$-variate polynomial $f_i$ based on the wiring predicate.
There is some flexibility in choosing the specific
polynomial $f_i$ to use. 
This is because the definition of $f_i$ involves a low-degree
extension of the circuit's wiring predicate, and there are many such
low-degree extensions to choose from. 

A polynomial is said to be \emph{multilinear} if it has degree
at most one in each variable. The results in this section rely
critically on the observation that the honest prover's computation in the protocol of \cite{muggles}
can be greatly simplified if we use the multilinear extension of the circuit's wiring predicate.\footnote{There are other reasons why using the multilinear extension is desirable. For example, the 
communication cost of the protocol is proportional to the degree of the extension used.}
Details of this observation follow.

As already mentioned, at iteration $i$ of the protocol of \cite{muggles}, the sum-check protocol
is applied to the $3v$-variate polynomial $f_i$. In the $j$'th round of this sum-check protocol, 
$\P$ is required to send the \emph{univariate} polynomial
\begin{align*} g_j(X_j) &\!=\! \sum_{ (x_{j+1}, \dots, x_{3v})\in\{0,1\}^{3v-j}}
f_i(r^{(i)}_1, \dots, r^{(i)}_{j-1}, X_j, x_{j+1}, \dots, x_{3v}).\end{align*}
The sum defining $g_j$ involves as many as $n^3$ terms, and thus
a naive implementation 
of $\P$ would require $\Omega(n^3)$ time per iteration of the protocol.
However, we observe that if the multilinear extension of the circuit's wiring predicate
is used in the definition of $f_i$, then each gate at layer $i-1$ contributes to exactly
one term in the 
sum defining $g_j$, as does each gate at layer $i$. 
Thus, the polynomial $g_j$
can be computed with a \emph{single} pass over the gates at layer $i-1$,
and a single pass over 
the gates at layer $i$. 
As the sum-check protocol requires $O(v) = O(\log S(n))$ messages for each layer of
the circuit, 
$\P$ requires logarithmically many passes over each layer of the
circuit in total. 

A complication in applying the above observation is that $\V$ must
process the circuit in order to pull out information about its
structure necessary to check the validity of $\P$'s
messages. 
Specifically, each application of the sum-check protocol requires $\V$ to evaluate the multilinear extension
of the wiring predicate of the circuit at a random point. 
Theorem \ref{OPT1}
follows from the fact that for any log-space uniform circuit,
$\V$ can evaluate the multilinear extension of the wiring predicate at
any point using space $O(\log S(n))$.  
We present detailed proofs and discussions of the following theorems
in Appendix~\ref{fullapp:muggles}. 


\begin{theorem}
\label{OPT1}
For any log-space uniform circuit of size $S(n)$,
$\P$ requires $O(S(n) \log S(n))$ time 
to implement the protocol of Theorem~\ref{thm:muggles}
over the entire execution, and $\V$ requires space $O(\log S(n))$.
\end{theorem}

Moreover, because the circuit's wiring predicate is independent of the input, we can separate $\V$'s computation into an offline non-interactive preprocessing phase, which occurs before the data stream is seen, and an online interactive phase which occurs after both $\P$ and $\V$ have seen the input. This is similar to \cite[Theorem 4]{muggles}, and ensures that $\V$ is space-efficient (but may require time $\poly(S(n))$) during the offline phase, and that $\V$ is \emph{both} time- and space-efficient in the online interactive phase. In order to determine which circuit to use, $\V$ does need to know (an upper bound on) the length of the input during the preprocessing phase.

\begin{theorem}
\label{thm:online}
For any log-space uniform circuit of size $S(n)$,
$\P$ requires $O(S(n) \log S(n))$ time 
to implement the protocol of Theorem~\ref{thm:muggles}
over the entire execution. $\V$ requires space $O(d(n)\log S(n))$ and time
$O(\poly(S(n)))$ in a non-interactive, data-independent preprocessing
phase, and only requires space
$O(d(n)\log S(n))$ and time $O(n \log n + d(n)\log S(n))$ in an online interactive phase, where the $O(n \log n)$ term is due to the 
time required to evaluate the low-degree extension
of the input at a point.
\end{theorem}

Finally, Theorem \ref{OPT2} follows by assuming $\V$ can evaluate the multilinear extension of the wiring predicate quickly. A formal statement of Theorem \ref{OPT2} is in Appendix \ref{fullapp:muggles}.
We believe that the hypothesis of Theorem \ref{OPT2} is extremely mild, and we discuss this point at length in Appendix \ref{fullapp:muggles}, identifying a diverse array of circuits to which Theorem \ref{OPT2} applies. 
Moreover, the solutions we adopt in our circuit-checking experiments for \sjs, \distinct, and \pmw\ correspond to Theorem \ref{OPT2}, and are both space- and 
time-efficient for the verifier. 

\begin{theorem}
\label{OPT2}
(informal) Let $C$ be any log-space uniform circuit of size $S(n)$ and depth $d(n)$, and assume there exists a
$O(\log S(n))$-space, $\poly(\log S(n))$-time algorithm for evaluating the multilinear extension
of $C$'s wiring predicate at a point. Then in order to to implement the protocol of Theorem~\ref{thm:muggles} applied to $C$,
$\P$ requires $O(S(n) \log S(n))$ time, and $\V$ requires space
$O(\log S(n))$ and time $O(n \log n + d(n)\poly(\log S(n)))$,
where the $O(n \log n)$ term is due to the 
time required to evaluate the low-degree extension
of the input at a point.
\end{theorem}

}

\subsection{Circuit Design Issues}
\label{app:muggles}

The protocol of \cite{muggles} is described for arithmetic
circuits with addition ($+$) and multiplication gates ($\times$).
This is sufficient to prove the power of this system, since any
efficiently computable boolean function on boolean inputs
can be computed by an (asymptotically) small arithmetic circuit.
Typically such arithmetic circuits are obtained by constructing a 
\emph{boolean} circuit (with AND, OR, and NOT gates) for the function,
and then ``arithmetizing'' the circuit \cite[Chapter 8]{arorabarak}.
However, we strive not just for asymptotic efficiency, but genuine
practicality, and the factors involved can grow quite quickly: every
layer of (arithmetic) gates in the circuit adds $3v$ rounds of
interaction to the protocol.  
Hence, we further explore 
optimizations and implementation issues. 

\para{Extended Gates.}
The circuit checking protocol of \cite{muggles} 
can be extended with any gates that  
 compute low-degree polynomial functions of their inputs. 
If $g$ is a polynomial of degree $j$, we can use gates 
computing $g(x)$; this
increases the communication complexity in each round of the protocol
by at most $j-2$ words, as $\P$ must send a degree-$j$
polynomial, rather than a degree-2 polynomial. 

\graham{The low-depth circuits we use to compute functions of interest
(specifically, \distinct\ and \pmw) 
make use of the function}
 $f(x) = x^{p-1}$.
Using only $+$ and $\times$ gates, they require depth about
$\log_2 p$. 
If we also use gates computing $g(x, y)=x^jy^j$ for a small $j$, 
we can reduce the depth of the circuits to
about $\log_{2j} p$; as the number of rounds in the protocol of
\cite{muggles} depends linearly on the depth of the circuit, this
reduces the number of rounds by a factor of about
$\log_2 p/\log_{2j} p = 1/\log_{2j} 2$. 
At the same time this increases the communication 
cost of each round by a factor of (at most) $j-2$. 
We can optimize the choice of $j$. 
In our experiments, we use $j=4$ (so $g(x,x)$ is $x^{8}$)
and $j=8$ ($g(x,x)=x^{16}$) to 
simultaneously reduce the number of messages by a factor of 3, 
and the communication cost and prover
runtime by significant factors as well.

Another optimization is possible. 
All four specific problems we consider, 
\sjs, \distinct, \pmw, and \matvect, 
eventually compute the sum of a large number of values. 
Let $f$ be the low-degree extension of the values being summed.
For functions of this form, 
$\V$ can use a single \emph{sum-check protocol} \cite[Chapter 8]{arorabarak}
to reduce the computation of the sum to computing $f(r)$ for a random
point $r$. 
$\V$ can then use the protocol of \cite{muggles} 
to delegate computation of $f(r)$ to $\P$. 
Conceptually, this optimization corresponds to replacing a binary tree
of addition gates in an arithmetic circuit $C$ with a single 
$\oplus$ gate with large fan-in, which sums all its inputs. 
This optimization can reduce the communication cost and the number of
messages required by the protocol.

\para{General Circuit Design.}  The circuit checking approach can be
combined with existing compilers, such as that in the Fairplay system \cite{fairplay},
that take as input a program in a high-level programming language and
output a corresponding boolean circuit. This boolean circuit can then
be arithmetized and ``verified'' by our implementation; this yields a
full-fledged system implementing statistically-secure verifiable
computation.  However, this system is 
likely to remain impractical even though the prover $\P$ can be made to run in
time linear in the size of the arithmetic circuit.  For example, in most hardware, one can compute
the sum of two 32-bit integers $x$ and $y$ with a single instruction.
However, when encoding this operation into a boolean circuit, it is
unclear how to do this with depth less than $32$.  At $3 \log n$
rounds per circuit layer, for reasonable parameters, single
additions can turn into thousands of rounds.  

The protocols in Section \ref{sec:effp} avoid this by avoiding boolean
circuits, and instead view the input directly as elements over \Fp.
For example, if the input is an array of 32-bit integers, then we view
each element of the array as a value of \Fp, and calculating the sum
of two integers requires a single depth-1 addition gate, rather than a
depth-32 boolean circuit.  However, this approach seems to severely
limit the functionality that can be implemented.  For instance, we
know of no compact arithmetic circuit to test whether $x
> y$ when viewing $x$ and $y$ as elements of \Fp.  Indeed, if such a 
circuit for this function existed, we would
obtain substantially improved protocols for \distinct\ and \pmw.

This polylogarithmic blowup in circuit depth compared to input size
appears inherent in any construction that encodes
computations as arithmetic circuits. 
Therefore, the development of general purpose protocols that avoid this
representation remains an important direction for future work.

\subsection{Efficient Protocols For Specific Problems}
\label{sec:effp}
We obtain \emph{interactive} protocols for our problems of interest by applying
Theorem \ref{thm:muggles} to carefully chosen arithmetic circuits.
These are circuits where each gate executes a simple arithmetic
operation on its inputs, such as addition, subtraction, or multiplication. 
For the first three problems, there exist specialized protocols;
our purpose in describing these protocols here is to explore how the
general construction performs when applied to specific functions of
high interest.
However, for \pmw, the protocol we describe here is the first of its kind.

For each problem, we describe a circuit which 
exploits the arithmetic structure of the finite field over
which they are defined.  For the latter three problems, 
this involves an interesting use of Fermat's Little Theorem.
These circuits lend themselves to extensions of
the basic protocol of \cite{muggles} that achieve quantitative
improvements in all costs; we demonstrate the extent of these
improvements in Section \ref{sec:experiments}.

\para{Protocol for \sjs}:
The arithmetic circuit for \sjs\ is quite straightforward: 
the first level computes the square of input values, then subsequent
levels sum these up pairwise to obtain the sum of all squared values.  
The total depth $d$ is $O(\log n)$. 
This implies a $O(\log^2 n)$ message
$(\log^2 n, \log^2 n)$ protocol (as per Definition \ref{def:protocol}).

\para{Protocol for \distinct}: 
We describe a succinct arithmetic circuit over $\mathbb{F}_p$ that
computes \distinct. 
When $p$ is a prime larger than $n$, Fermat's Little Theorem (FLT)
implies that for
$x \in \Fp$, $x^{p-1}=1$ if and only if $x \neq 0$. 
Consider the circuit that, for each coordinate $i$ of the input vector $a$,
computes each $a_i^{p-1}$ via $O(\log p)$ multiplications, 
and then sums the results. 
This circuit has total size $O(n \log p)$ and depth $O(\log p)$.  
Applying the protocol of \cite{muggles} to this circuit, 
we obtain a $(\log n \log p, \log n)$ protocol where $\P$ runs in time
$O(n \log n \log p)$.

\para{Protocol for \matvect}:
The first level of the circuit computes $A_{ij}\mathbf{x}_i$ for all
$i, j$, and subsequent levels sum these to obtain $\sum_{j} A_{ij}
\mathbf{x}_i$. 
Then we use FLT to ensure that $\sum_{j} A_{ij} \mathbf{x}_i = b_i$ 
for all $i$, via 
\[ \sum_i \left(\left(\sum_{j} A_{ij} \mathbf{x}_i\right) - b_i\right)^{p-1} . \]
The input is as claimed if this final output of the circuit is $0$
(i.e. it counts the number of entries of $\mathbf{b}$ that are incorrect). 
This circuit has depth $O(\log p)$ and and size $O(n^2 \log p)$, and we
therefore obtain an $(n+ \log p \log n, \log n)$ protocol requiring
$O(\log p \log n)$-rounds, 
where $\P$ runs in time $O(n^2 \log p \log n)$.

\para{Protocol for \pmw}: 
\eat{We first present a protocol for pattern-matching without wildcards. 
Let $p$ be a prime larger than $n$. 
Given a stream representing text
$T = (t_0, \dots , t_{n-1}) \in [n]^n$ and pattern $P= (p_0,
\dots , p_{q-1})\in[n]^q$, our circuit computes $n-q$ vectors\\
$\mathbf{v}^{(0)}, \dots, \mathbf{v}^{(n-q-1)} \in \Fp^q$,
where 
\[\mathbf{v}^{(i)}=(t_i-p_0, t_{i+1} - p_1, \dots,
t_{i+q-1}-p_{i+q-1});\] 
this requires just one layer with $nq$ gates.  
Then for each vector $\mathbf{v}^{(i)}$, the circuit computes the $p-1$'th
power of each entry, sums these, subtracts the results from $q$, and
takes the $p-1$'th power of the results.  By Fermat's Little Theorem,
this gives an indicator vector $I$ for where the pattern (does not)
occur in $T$.  Finally, the circuit sums the entries in the indicator
vector $I$ to obtain the output value $y$.  The number of occurrences
of the pattern equals $n-q-y$.  
This circuit has size $O(nq \log p)$ and depth $O(\log p)$.
}
To handle wildcards in both $T$ (of length $n$) and $P$ (of length $q$),
we replace each occurrence of the wildcard symbol with $0$;
\cite{wildcards} notes that the pattern
occurs at location $i$ of $T$ if and only if
\[ I_i:=\sum_{j=0}^{q-1} t_{i+j}p_{j}(t_{i+j}-p_{j})^2 \neq 0.\] 
Thus, by FLT, it suffices to compute $\sum_{i=0}^n I_i^{p-1}$, which
can be done naively by an arithmetic circuit of size $O(nq + n\log p)$ and depth
$O(\log p + \log q)$. 
We obtain a $( \log n \log p, \log n)$ protocol where 
$\P$ runs in time $O(n\log n (q + \log p))$. 

For large $q$, a further optimization is possible: the 
vector $I$ can be written as the sum of a constant number of circular
convolutions.
Such convolutions can be computed efficiently using Fourier techniques
in time $O(n \log q)$ and, importantly, appropriate FFT and inverse
FFT operations can be implemented via arithmetic circuits. 
Thus, for $q$ larger than $\log p$, we can reduce the circuit size
 (and hence $\P$'s runtime) in this way, rather than by naively
computing each entry of $I$ independently.


\section{Multi-Round Protocols via Linearization}\label{sec:distinct}
In this section, we show how the technique of {\em linearization} can improve
upon the general approach of Section \ref{sec:pattern} for some
important functions.  
Specifically, this technique can be applied to multi-round protocols which would
otherwise require polynomials of very 
high degree to be communicated. 
We show this in the context of new multi-round protocols for
\distinct\ and \pmw\, and we later empirically observe that our new protocol achieves a speed up
of
two orders of magnitude over existing protocols for \distinct, as well as an order of magnitude improvement
in communication cost.

\eat{
Since the seminal paper of Flajolet and Martin nearly thirty years ago
\cite{fm85}, the {\sc Distinct} problem has been the subject of a long
line of research in the streaming community.
This recently led to the
space-optimal approximation algorithm by Kane et al.~\cite{kane}.  
The problem has applications in a number of areas, from
data mining to query optimization~\cite{kane}.  
Here, we give a new protocol for computing $F_0$ 
\emph{exactly} in our multi-round model.}

Existing approaches for \distinct\ in the 
multi-round setting are based on generalizations of the multi-round
protocol for \sjs~\cite{pods}. 
As described in \cite{pods}, directly applying this approach is
problematic: the central function 
in \distinct\ maps non-zero frequencies to 1 while keeping zero
frequencies as zero. 
Expressed as a polynomial, this function has degree $m$ (an upper
bound on the frequency of any item), which
translates into a factor of $m$ in the communication required and the
time cost of $\P$. 
However, this cost can be reduced to $F_{\infty}$, where $F_{\infty}$
denotes the maximum number of times any item appears in the stream. 
Further, if both $\P$ and $\V$ keep a buffer of $b$ input items, they can
eliminate duplicate items within the buffer, and so ensure that 
$F_{\infty} \leq m/b$. 
This leads to an $O(\log n)$ message, 
$(\log n, F_{\infty}\log n)$ multi-round protocol 
with $\P$'s runtime being $O(F_{\infty}^2 n \log n)$ \cite{pods}. 
This protocol trades off increased 
communication for a quadratic improvement in the number of
rounds of communication required compared to the protocol
outlined in Section~\ref{sec:effp} above. 

\subsection{Linearization Set-up}
\label{sec:linear}
In this section we describe a new multi-round protocol for
\distinct, and later explain how it can be modified for \pmw.
This protocol has similar asymptotic costs as that obtained in Section
\ref{sec:effp}, 
but in practice achieves close to two orders of magnitude improvement
in $\P$'s runtime. 
The core idea is to represent the data as a large binary vector
indicating when each item occurs in the stream. 
The protocol simulates progressively merging time ranges together to
indicate which items occurred within the ranges.  
Directly verifying this computation would hit the same roadblock
indicated above: using polynomials to check this would result in
polynomials of high degree, dominating the cost.
So we use a ``linearization'' technique, which ensures that the degree
of the polynomials required stays low, at the cost of more
rounds of interaction. 
This uses ideas of Shen \cite{shen92} as presented in \cite[Chapter
  8]{arorabarak}.  
 
As usual, we work over a
finite field with $p$ elements, $\mathbb{F}_p$.
The input implicitly defines an $n \times m$ matrix $A$ 
such that 
$A_{i, j}=1$ if the $j$'th item of the stream equals $i$,
and $A_{i,j}=0$ otherwise. 

\para{Working over the Boolean Hypercube.}
A key first step is to define an indexing structure based on the
$d$-dimensional Boolean hypercube, so every input point is indexed by
a $d$ bit binary string, which is the (binary) concatenation of 
a $\log n$ bit string $i$ and a $\log m$ bit string $j$. 
We view $A$ as a function from 
$\{0, 1\}^d$ to $\{0,1\}$ via $(x_1, \dots, x_d) \mapsto 
A_{(x_1, \dots, x_d)}$.  
Let $f$ be the unique multilinear
polynomial in $d$ variables such that $f(x_1, \dots, x_d)
= A_{(x_1, \dots, x_d)}$ for all $(x_1, \dots, x_d) \in \{0,1\}^d$, i.e. 
$f$ is the \emph{multilinear extension} of the function on $\{0,1\}^d$ implied by $A$. 



The only information that the verifier $\V$ needs to keep track of is
the value of $f$ at a random point. 
That is, $\V$ chooses a random vector $\r = (r_1, \dots, r_d) \in
\mathbb{F}_p^d$.
It is efficient for $\V$ to compute $f(\r)$ 
as $\V$ observes the stream which defines $A$ (and hence
$f$): 
  when the $j$'th update is item $i$, this translates to the
 vector $\v=(i, j) \in \{0, 1\}^d$.
The necessary update is of the form
$f(\r) \leftarrow f(\r) + \chi_{\v}(\r)$,
where $\chi_{\v}$ is the unique polynomial that is 1 at $\v$ and 
0 everywhere else in $\{0,1\}^d$. 
For this, 
  $\V$ only needs to store $\r$ and the current value of $f(\r)$. 

\eat{
We can define $f$ in a constructive manner as follows.  Let $\x = (x_1,\dots,x_d) \in [p]^d$.  First,
note that the polynomial corresponding to a function which is $1$ at
location $\v = (v_1,\dots, v_d) \in \{0, 1\}^d$ and zero elsewhere in
$\{0, 1\}^d$ is 
\[ \chi_\v(\x) = \prod_{j=1}^{d} \chi_{v_j} (x_j) \]
where
$\chi_k(x_j)$ is the Lagrange polynomial given by 
$1-x_j$ if $k=0$, and $x_j$ if $k=1$,
which has the property that $\chi_k(x_j) = 1$ if $x_j=k$ and $0$ for $x_j = 1-k$.
We then define \[f(\x) = \sum_{\v \in \{0,1\}^d} f(\v) \chi_\v(\x).\] One can
easily verify that such an $f$ agrees with $f$ at all points on
the $d$-dimensional hypercube, as required.

\para{Verifier's computation.}
In our protocol, the only information the verifier need extract from
the stream is the value of $f$ at a random point. 
That is, $\V$ chooses a random vector $\r = (r_1, \dots, r_d) \in
\mathbb{F}_p^d$, and while observing  
the stream, $\V$ computes $f(\r)$ in an iterative fashion
as follows. 
When the $j$'th update is observed  to be item $i$, this is associated
with vector $\v=(i, j) \in \{0, 1\}^d$.
Therefore, the necessary update is of the form
$f(\r) \leftarrow f(\r) + \chi_{\v}(\r)$, and
$\V$ need only store $\r$ and $f(\r)$ while processing the input.
}
\vspace*{-1mm}
\medskip
\noindent \textbf{Linearization and Arithmetized Boolean Operators.}
We use three operators $\amalg$, $\Pi$ and $L$ on polynomials $g$, defined as follows:
\begin{align*} 
\amalg_k g(X_1, \dots, X_k) = &  
  g(X_1, \dots, X_{k-1}, 0) + g(X_1, \dots, X_{k-1}, 1) \\
  - & g(X_1, \dots, X_{k-1}, 0) \cdot g(X_1, \dots, X_{k-1}, 1). \\[6pt]
\Pi_k g(X_1, \dots, X_k) =\ &
  g(X_1, \dots, X_{k-1}, 0) \cdot g(X_1, \dots, X_{k-1}, 1).\\[6pt]
L_i g(X_1 , \dots , X_k) = &
  X_i \cdot g(X_1 , \dots, X_{i-1}, 1, X_{i+1}, \cdots , X_k)\\
   + & (1- X_i) \cdot g(X_1, \dots , X_{i-1}, 0, X_i, \dots , X_k).
\end{align*}

$\amalg$ and $\Pi$ generalize the familiar ``OR'' and ``AND''
operators, respectively. 
Thus, if $g$ is a $k$-variate polynomial of degree at most $j$ in each
variable, $\amalg_k(g)$ and $\Pi_k(g)$ are $k-1$-variate polynomials  
of degree at most $2j$ in each variable. 
They generalize Boolean operators in the sense that if
$g(X_1, \ldots, X_{k-1}, 0) = x$ and $g(X_1, \ldots, X_{k-1}, 1)=y$,
and $x, y$ are both 0 or 1, 
then 
\[
\begin{array}{rl}
(\amalg_k g)(X_1, \ldots X_k)=1 & \text{ iff $x=1$ or $y=1$, }\\[3pt]
\text{and }  (\Pi_k g)(X_1, \ldots X_k)=1 & \text{ iff $x=1$ and $y=1$}.
\end{array}
\] 
$L$ is a {\em linearization} operator.
If $g$ is a $k$-variate polynomial, $L_i(g)$ 
is a $k$-variate polynomial that is linear in variable $X_i$.
$L_i$ operations are used to control the degree of the polynomials
that arise throughout the execution of our protocol.
Since $x^j=x$ for all $j \geq 1, x\in\{0,1\}$, $L_i(g)$ agrees with
$g(\cdot)$ on all values in $\{0, 1\}^k$.

Throughout, when applying a sequence of operations to a polynomial to
obtain a new one, the operations are applied ``right-to-left''. 
For example, we write the $k-1$ variate polynomial 
\[(L_1 (L_2 \dots
(L_{k-1} (\amalg_k g)))) \text{ as } L_1 L_2 \dots L_{k-1} \amalg_k g .\]

\para{Rewriting \distinct\ and \pmw}.
For \sjs\ and \matvect\ there is little need for
linearization: the polynomials generated remain of low-degree, so the
multi-round protocols described in \cite{pods,graphstream} already suffice. 
But linearization can help with \distinct\ and \pmw.

Thinking of the input as a matrix $A$ as defined above, we can compute
\distinct\ by repeatedly taking the columnwise-OR of adjacent column pairs 
to end up with a vector which indicates whether item $i$ appeared in
the stream, then repeatedly summing adjacent entries to get the
number of distinct elements. When representing these operations as polynomials,
\graham{we make additional use of linearization operations} to control the degree of the polynomials that arise.
Using the properties of the operations $\amalg$ and $L_i$ described
above and rewriting in terms of the hypercube, it can be seen that 
\begin{align*}
 F_0(\a) = & \sum_{x_1=0}^1 \dots \sum_{x_{_k}=0}^1 L_{k_1} L_{k_1-1} \dots
 L_1 \amalg_{k_1+1} & L_{k_1+1}
 L_{k_1} \dots L_{1}
  \amalg_{k_1+2} \dots L_{d-1} L_{d-2} \dots L_1 \amalg_{d} f 
\tag{2}   \label{F0eq} \end{align*}
because this expression only involves variables and values in $\{0, 1\}$. 
The size of this expression is $\frac{3d^2+3d}{2}=O(\log^2n)$.  

The case for \pmw\ is similar.
Assume for now that the pattern length $q$ is a power of two (if not,
it can be padded with trailing wildcards).  
We now consider the input to define a matrix $A$ of size 
$2n \times qn$, such that 
$A_{2i,qj+k(q-1)} = 1$ if the $j$'th item of the stream equals $i$,
for all $0 \leq k \leq q-1$,  and 
$A_{2i-1,qj + 2k}=1$ if the $k$'th character of the pattern equals $i$,
for all $0 \leq j \leq n-1$. 
Wildcards in the pattern or the text are treated as occurrences of all
characters in the alphabet at that location. 
The problem is solved over this matrix $A$ by first taking the column-wise
``AND'' of adjacent columns: this leaves 1 where a text character
matches a pattern for a certain offset. 
We then take column-wise ``OR''s of adjacent columns $\log n$ times:
this collapses the alphabet. 
Taking row-wise ``AND''s of adjacent rows $\log q$ times leaves an
indicator vector whose $i$th entry is 1 iff the pattern occurs at
location $i$ in the text. 
Summing the entries in this vector provides the required answer.
Using linearization to bound the degree of $\amalg$ and $\Pi$
operators, we again obtain an expression of size $O(\log^2 n)$. 

\subsection{Protocols Using Linearization}
Given an expression in the form of \eqref{F0eq}, we 
now give an inductive description of the protocol.
Conceptually, each round we ask the prover to ``strip off'' the left-most
remaining operation in the expression.
In the process, we reduce a claim by $\P$ about the old expression
to a claim about the new, shorter expression. 
Eventually, $\V$ is left with a claim about the value of 
$f$ at a random point (specifically, at $\r$), 
which $\V$ can check against her independent evaluation of $f(\r)$.

More specifically, suppose for some polynomial $g(X_1 , \dots , X_j)$, the prover 
can convince the verifier that $g(a_1, a_2, \dots, a_j)=C$
 with probability 1 for any $(a_1, a_2 , \dots, a_j, C)$ where 
this is true, and probability less than $\epsilon$ when it is false. 
Let $U(X_1 , X_2 , \dots , X_l)$ be any polynomial on $l$ variables obtained as 
\[U(X_1 , X_2 , \dots , X_l ) = \mathcal{O} g(X_1 , \dots , X_j),\] 
where $\mathcal{O}$ is one of 
$\sum_{x_i=0}^1$, $\Pi_{x_i=0}^1$, $\amalg_{x_i=0}^1$ or $L_i$ for some variable $i$. 
(Thus $l$ is $j-1$ in the first three  cases and $j$ in the last). 
Let $m$ be an upper bound (known to the verifier) on the degree 
of $U$ with respect to $X_i$. In our case, $m \leq 2$ because of the inclusion of $L_i$ operations in between every $\amalg$ and $\Pi$ operation. 
We show how $\P$ can convince $\V$ that $U (a_1 , a_2 , \dots , a_l ) = C'$ with probability 1 for any $(a_1 , a_2 , \dots , a_j , C')$ for which it 
is true and with probability at most $\epsilon + d/p$ when it is false. 
By renaming variables if necessary, assume $i = 1$. 
The verifier's check is as follows.

\para{Case 1:} $\mathcal{O} = \sum_{x_1 = 0}^1$. 
$\P$ provides a degree-1 polynomial $s(X_1)$ that is supposed to be 
$g(X_1 , a_2 , \dots , a_j)$. 
$\V$ checks if $s(0) + s(1) = C'$. 
If not, $\V$ rejects. If so, $\V$ picks a random value 
$a \in \mathbb{F}_p$ and asks $\P$ to prove $s(a) = g(a, a_2 , \dots ,
a_j)$. 
If it is one of the final
$d$ rounds, $\V$ chooses $a$ to be the corresponding entry of $\r$. 

\para{Case 2:}
$\mathcal{O} = \amalg_{x_1=0}^1 X$ or $\mathcal{O} = \Pi_{x_1=1}^1 X$. 
We do the same as in Case 1, but replace $s(0) + s(1)$ with
 $s(0)+s(1) - s(0)s(1)$ in the case of $\amalg$,
or $s(0)s(1)$ in the case of $\Pi$. 

\para{Case 3:} $\mathcal{O} = L_1$. 
$\P$ wishes to prove that $U (a_1 , a_2 , \dots , a_k ) = C'$. 
$\P$ provides a degree-2 polynomial $s(X_1)$ that is supposed to be 
$g(X_1 , a_2 , \dots , a_k)$. 
We refer to this as ``unbinding the variable'' because previously 
$X_1$ was ``bound'' to value $a_1$, but now $X_1$ is free. 
$\V$ checks that $a_1 s(0) + (1 - a_1 )s(1) = C'$. 
If not, $\V$ rejects. 
If so, $\V$ picks random  $a \in \mathbb{F}_p$ and asks $\P$ to prove 
$s(a) = g(a, a_2 , \dots , a_k)$ 
(or if it is the final round, $\V$ simply checks that $s(a)=f(\r))$.

\medskip
The proof of correctness follows by using the observation that if 
$s(X_1)$ is not the right polynomial, then with probability $1-m/p$, 
$\P$ must prove an incorrect statement at the next round
(this is an instance of Schwartz-Zippel polynomial equality testing
  procedure \cite{schwartz80:_fast}). 
The total probability of error is given by a union bound on the
probabilities in each round, $O(\log^2 n/p)$.
  
\para{Analysis of protocol costs.} 
Recall that both \distinct\ and \pmw\ can be written as an expression
of size $O(\log^2 n)$ operators, where linearization bounds the degree
in any variable. 
Under the above procedure, 
the verifier need only store $\r$, $f(\r)$, the
current values of any ``bound'' variables, 
and the most recent value of $s(a)$. 
In total, this requires space $O(\log n)$. 
There are $O(\log^2 n)$ rounds, and in each round a polynomial 
of degree at most two is sent from $\P$ to $\V$. 
Such a polynomial can be represented with at most $3$ words, 
so the total communication is $O(\log^2 n)$. 
Hence we obtain ($\log^2 n, \log n)$-protocols for \distinct\ and \pmw.

As the stream is being processed the verifier has to update $f(\r)$. 
The updates are very simple, and  processing each update requires
$O(d) = O(\log n)$ time. 
There is a slight overhead in \pmw, where 
each update in the stream requires the verifier to
propagate $q$ updates to $f$ 
(assuming an upper bound on $q$ is fixed in advance), taking $O(q)$ time. 
However, it seems plausible that these costs could be optimized 
further.

The prover has to store a description of the 
stream, which can be done in space $O(n)$. 
The prover can be implemented to require $O(n \log^2 n)$
time: essentially, each round of the proof requires at most one pass
over the stream data to compute the required functions. 
For brevity, we omit a detailed description of the implementation,
\graham{the source code of which is available at \cite{code}.}

\begin{theorem}
\graham{For any function which can be written as a concatenation of
  $\log n$ (binary) operators drawn from $\sum, \Pi$ and $\amalg$ over inputs of size $n$,
there is a $\log^2 n$ round $(\log^2 n, \log n)$ protocol, where $\P$ takes time $O(n
\log^2 n)$, and $\V$ takes time $O(\log^2 n)$ to run the protocol,
having computing the LDE of the input.
}
\end{theorem}

\graham{Thus we can invoke this theorem for both \distinct\ and \pmw, 
obtaining $\log^2n$ round $(\log^2 n, \log n)$ protocols for both. }

\section{Experimental Evaluation}
\label{sec:experiments}

We performed a thorough experimental study to evaluate the potential
practical effectiveness of existing protocols and our new ones. 
We summarize our findings as follows.  

\begin{table*}[t]
\centering
\begin{tabular}{ | c | c | c | r@{.}l | c | c@{\hspace{6ex}} r@{.}l | c |}
\hline
Problem&Gates&  Size          & \multicolumn{2}{c|}{$\P$ time}
&Rounds& \multicolumn{3}{|c|}{Communication} & $\V$ time       \\  
               &                &  (gates) &
  \multicolumn{2}{|c|}{(s)}&                    &
  \multicolumn{3}{c|}{(KBs)} & (s)         \\
\hline
\sjs & $+,\times$          & 0.4M 
&   8&5    &986& &11&5  &   .01 \\
\sjs & $+,\times,\oplus$& 0.2M 
&   6&5    &118& &2&5 &   .01 \\
\distinct          &$+,\times$& 16M 
&  552&6 &3730& &87&4 & .01 \\
\distinct          &$+,\times,\mathbin{\char`\^}8$& 8.4M 
&  462&2 &1684&  &60&0 & .01 \\
\distinct         &$+,\times,\mathbin{\char`\^}16$& 6.4M 
& 457&4 &1399& &65&8  & .01 \\
\distinct          &$+,\times,\oplus$& 15.8M 
&  546&4 &3355& &78&4 & .01 \\
\distinct          &$+,\times,\mathbin{\char`\^}8,\oplus$& 8.2M 
&  432&6 &1310&  &51&0 & .01 \\
\distinct         &$+,\times,\mathbin{\char`\^}16,\oplus$& 6.2M 
& 441&2 &1024& &56&8  & .01 \\
\pmw             &$+,\times,\mathbin{\char`\^}8,\oplus$& 9.6M 
& 482&2 &1513& &56&1 & .01 \\
\hline
\end{tabular}
\caption{Circuit checking results with $n=2^{17}$.}
\label{tabmuggles}
\end{table*}

\begin{itemize}
\item
The costs of our implementation of
the general-purpose circuit-checking protocol described in
Section~\ref{sec:muggles} are
extremely attractive, with the exception of $\P$'s runtime. 
The prover takes minutes to operate on input of size around $10^{5}$: 
ideally, this would take seconds. 
The extensions we propose to the 
basic protocol of \cite{muggles} \graham{(such as extra types of gates)} 
result in significant
quantitative improvements for our benchmark problems. 
We are optimistic about the prospects for further enhancements and
parallelization to make practical general-purpose verification a
reality.  

\item
Fine-tuned protocols for specific problems can improve over the
general approach by several orders of magnitude. 
Specifically, we found that extremely practical \emph{non-interactive}
protocols processing hundreds of thousands of updates per second are achievable 
for a very large class of problems, but only by
using the methods described in Section \ref{sec:FFT}.  
We also found
that the linearization technique results in significantly improved
interactive protocols for \distinct\ when compared to the more general
circuit-checking approach.

\item
Finally, we demonstrate that the non-interactive protocols are
extremely amenable to parallelization, 
and we believe that this makes them an attractive option for practical use.
\end{itemize}

\edit{In all of our experiments, the verifier requires significantly less space than that required to 
solve the problem without a prover, and requires about the same time
as that required to solve the problem without a prover if given enough
fast memory to store the whole input. Indeed, we found that in all of
our protocols memory accesses are the speed bottleneck in both $\V$'s
computation and in the computation required to solve the problem
without a prover.  

Moreover, our circuit-checking results demonstrate that if we were to
run our implementation on problems requiring superlinear time to
solve, then $\V$ would save significant time as well as space
(compared to solving the problem without a prover). Indeed, except for
circuits with very high (i.e., linear) depth, $\V$'s runtime in our
circuit-checking implementation is grossly dominated by the time required to
perform an LDE computation via a single streaming pass over the
input. 
\graham{The verification time, excluding this cost, is essentially
  negligible.} 
}

\subsection{Implementation Details}
\label{sec:implementation}
All implementations were done in C++: we simulated the computations of
both parties, and measured the time and resources consumed by the
protocols.  Our programs were compiled with g++ using the -O3
optimization flag.  For the data, we generated synthetic streams in
which each item was picked uniformly at random from the universe, or
in which frequencies of each item were chosen uniformly at random in
the range $[0, 1000]$.  The choice of data does not 
affect the runtimes, which depend only on the amount of data and not
its content.  Similarly the security guarantees do not depend on
the data, but on the random choices of the verifier.  All computations
are over the field of size $p=2^{61}-1$, implying a very low
probability of the verifier being fooled by a dishonest prover.

We evaluated the protocols on a multi-core machine
with 64-bit AMD Opteron processors and 32 GB of memory available. 
Our scalability results primarily use a single core, but we also
show results for parallel operation. 
The large amount of memory allowed  us to experiment with universes of
size several billion, with the prover able to store
the full data in memory.
We measured the time for $\V$ to compute the check information
from the stream, for $\P$ to generate the proof, and for $\V$
to verify the proof. 
We also measured the space required by $\V$, and the
size of the proof provided by $\P$. 

\eat{
In order to quantify the increase in space, communication, and prover
time necessary in moving from multi-round to non-interactive
protocols, we compared the multi-round protocol for \sjs\ described in \cite{pods} to our improved implementation of the
non-interactive protocol from \cite{annotations} described in Section
\ref{sec:FFT}. We also compared the latter to the less efficient
implementation from \cite{pods}, in order to demonstrate the extent of
the improvement we were able to achieve.

Additionally, we compared our new multi-round protocol for \distinct\ to existing protocols,
and implemented the \matvect\
protocol of \cite{graphstream} using the FFT techniques of Section \ref{sec:FFT}. 
Finally, we developed an engineered implementation of the general purpose
construction of \cite{muggles}. 
}

\paragraph{Choice of Field Size}
While all the protocols we implemented work over arbitrary finite
fields, our choice of \Fp\ with $p=2^{61}-1$ proves ideal for engineering
practical protocols.  First, the field size is large enough to provide
a minuscule probability of error (which is proportional to $1/p$), but small enough that any field
element can be represented with a single 64-bit data type. By using
native types, we achieve a speedup of several factors.  
Second, reducing modulo $p$ can be done with a
bit shift, a bit-wise AND operation, and an addition \cite{Thorup:00}. 
We experienced a speedup of nearly an order of magnitude by 
switching to this specialized ``mod'' operation rather than using 
``\% p'' operation in C++. 
Finally, the use of this particular field allows us
to apply FFT techniques, as described in 
in Section \ref{sec:FFT} (recall $2^{61}-2$ has many small
prime factors).

\eat{
\begin{table}[!h]
\scalebox{.78}{
\begin{tabular}{ | c | c | c | c | c | c |}
\hline
Problem& Version & Circuit          &$\P$     & Messages & Communication        \\  
               &                & Size (gates) &time (s)&                    & (words)    \\
\hline
\sjs & basic            & 393215    &   8.48    & 986    & 1479   \\
\distinct          &  basic            &15990723 &  552.59 & 3730 & 11190\\
\distinct          &$\mathbin{\char`\^}8$ gates  & 8388566  &  462.21 & 1684 &  7691\\
\distinct         & $\mathbin{\char`\^}16$ gates & 6422496  & 457.37 & 1399 & 8427  \\
\hline
\end{tabular}
}
\caption{Experimental results with general-purpose implementation. Results are depicted
for streams with universe size $n=2^17$.}
\label{tabmuggles}
\end{table}
}

\para{Correctness of protocols.}
In the protocols we study, the verifier's checks of the prover's
claims are always very simple to implement: in many cases, 
each check takes a single line of code to ensure that the previous
message is consistent with the new message\footnote{Things are a
  little more complex in the case of circuit checking, as discussed in
  Section~\ref{app:muggles}, but not dramatically so.}.  
Consequently, it is not difficult to implement the verifier in a
bug-free manner, and once this is the case, the verifier's
implementation serves as an independent check on the prover's
implementation. 
This is because the verifier detects (with high probability) 
\emph{any} deviations from the prescribed protocol, and
in particular $\V$ detects deviations due to an incorrect prover.
Thus, we are confident in the correctness of our implementations.
More generally, this property can help in the testing and
debugging of future implementations. 

\newlength{\figwidth}
\setlength{\figwidth}{0.33\textwidth}
\begin{figure*}[t]
\subfigure[$\P$'s time for \sjs\ protocols]{
\includegraphics[width=\figwidth]{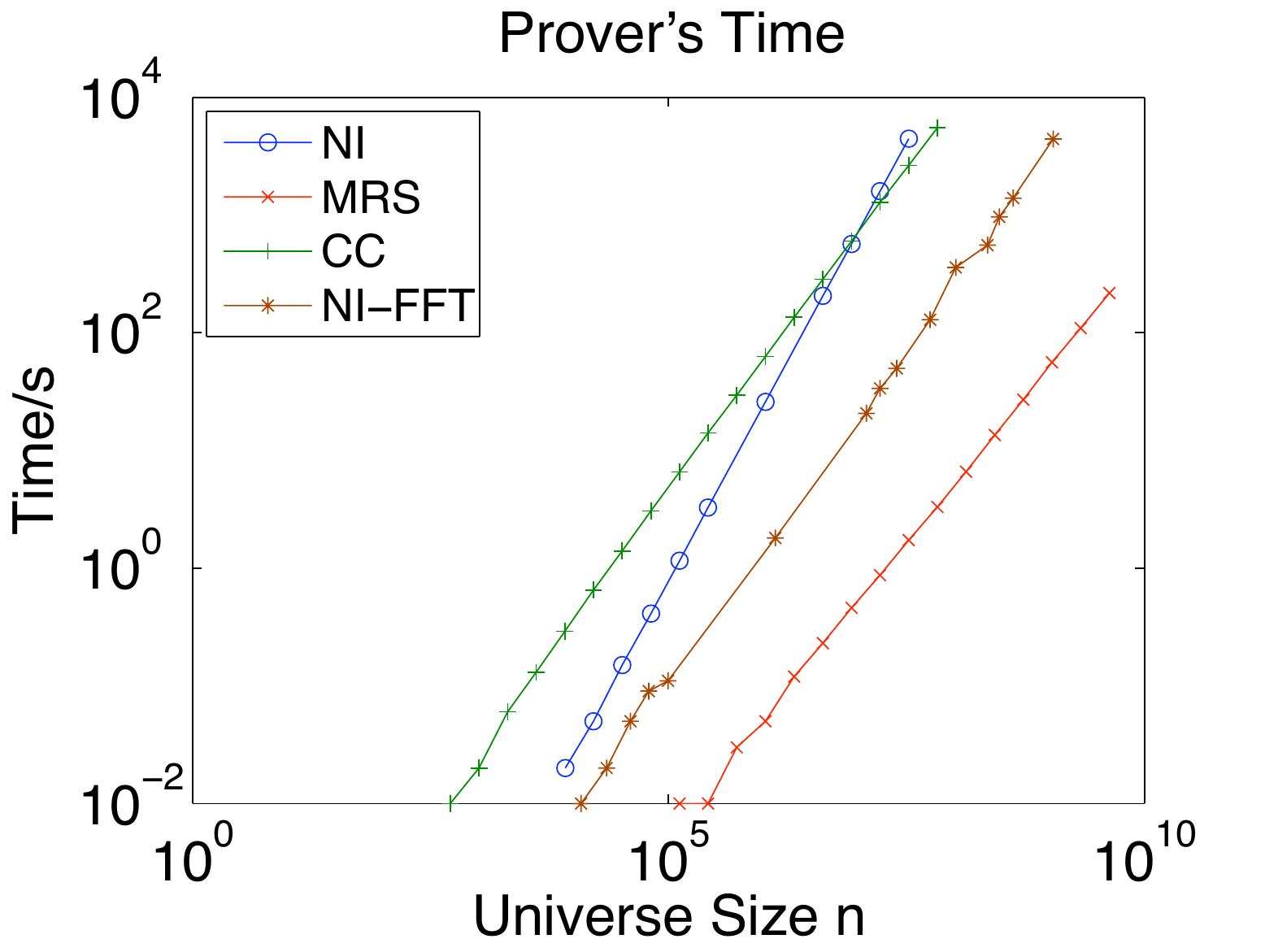}
\label{fig:timep}
}%
\subfigure[Verifier's time for \sjs\ protocols]{
\includegraphics[width=\figwidth]{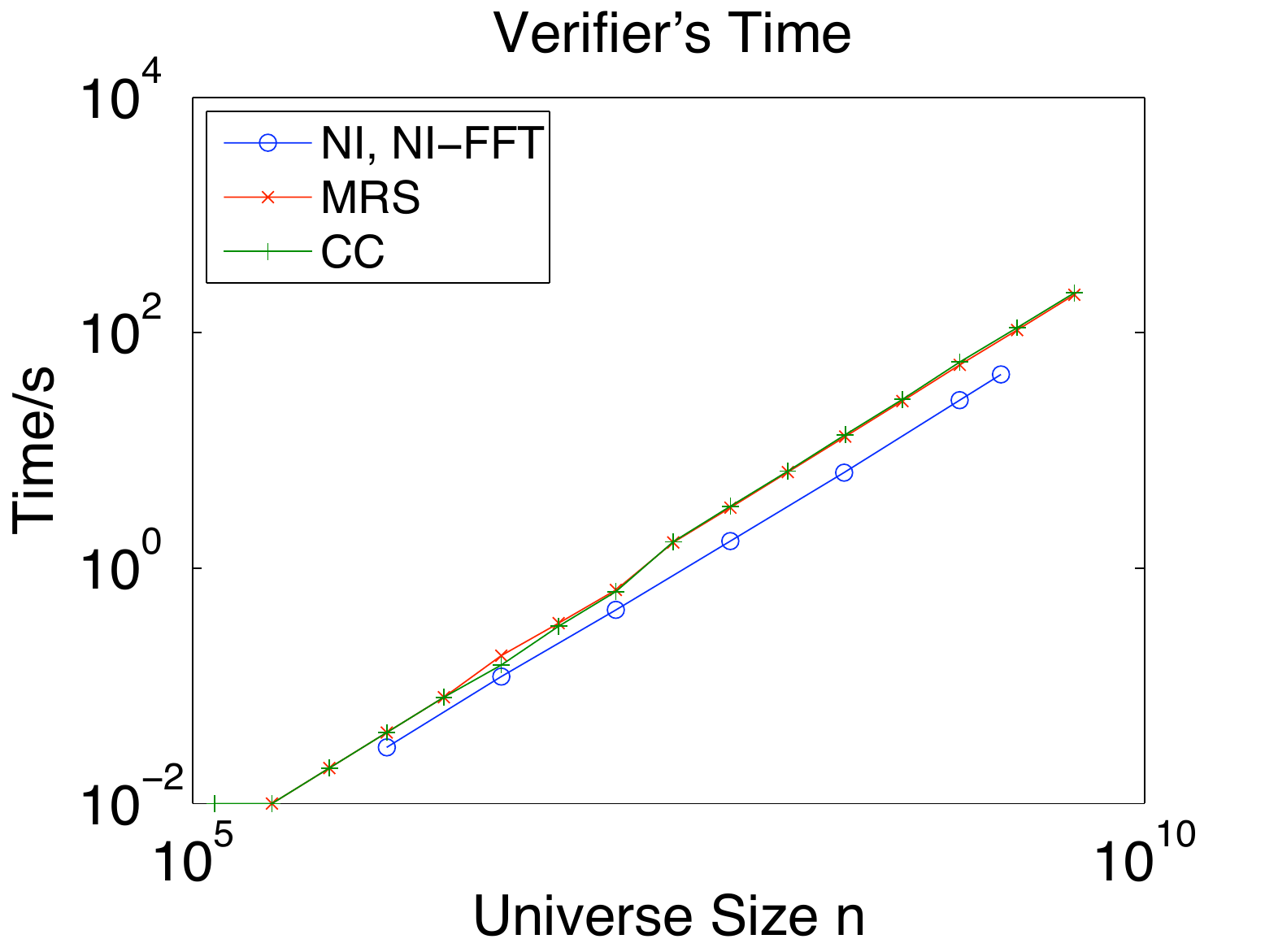}
\label{fig:timev}
}%
\subfigure[Space and communication costs for \sjs\ ]{
\includegraphics[width=\figwidth]{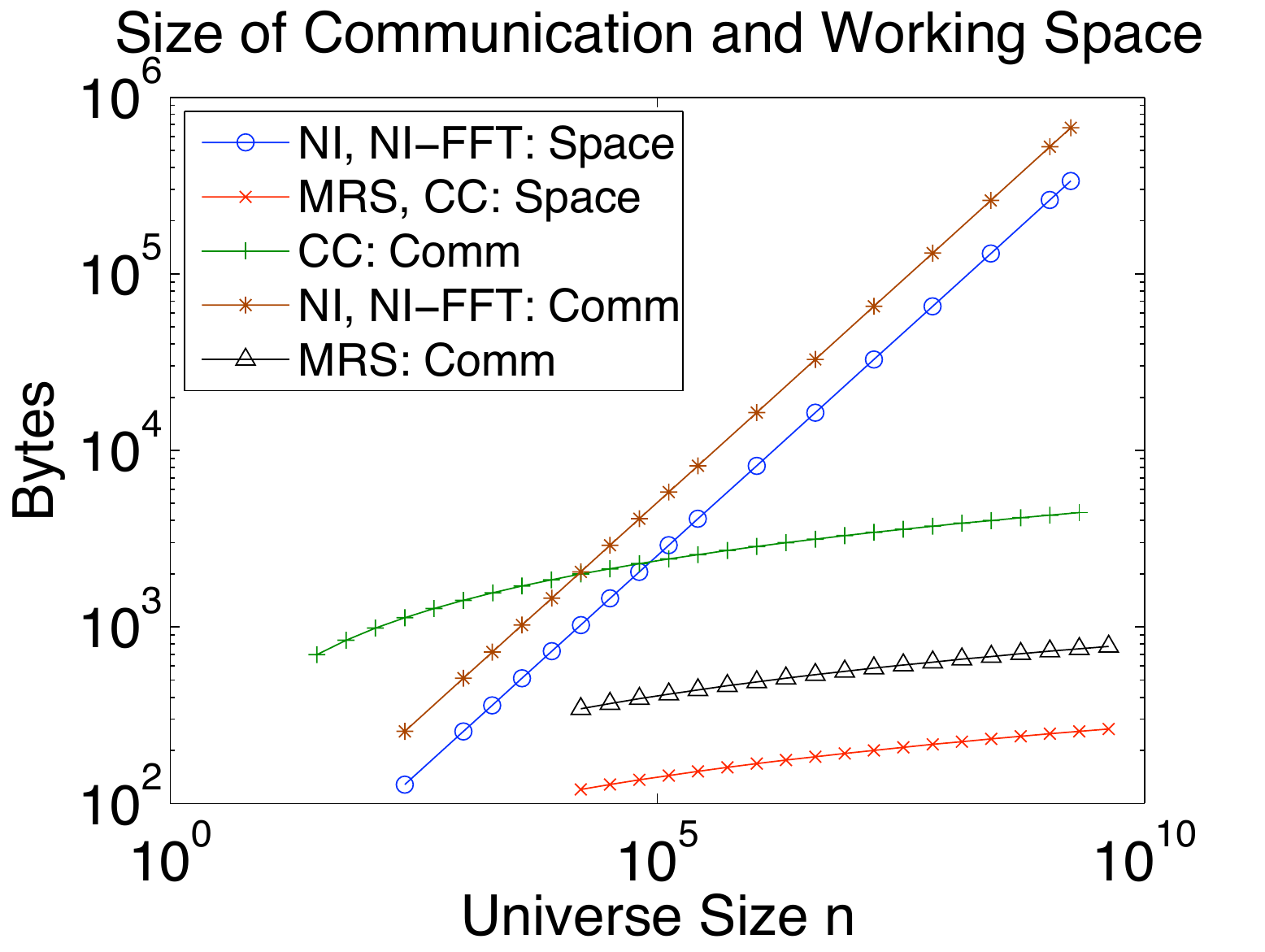}
\label{fig:space}
}
\caption{Experimental Results for both multi-round and non-interactive \sjs\ protocols.}
\label{fig:f2}
\end{figure*}

\subsection{Circuit Checking Protocols}
In our implementation of the circuit checking method described in 
Section \ref{sec:pattern}, we put significant effort into optimizing the
runtime of the prover, achieving an implementation for which $\P$
takes time nearly linear in the size of the circuit. 
Nonetheless, this cost remains the chief limitation of the
implementation.

We experimented with our implementation on circuits for three
of our functions of interest: \sjs, \distinct\ and \pmw. 
We leave circuits for \matvect\ to future work. 
Results are summarized in Table \ref{tabmuggles}. 
Throughout, when we refer to $\P$'s runtime in an interactive protocol,
we are referring  to the total time
\emph{over all rounds} of the protocol. 
The speed per gate can be very high: 
$\P$ processed circuits with tens
of millions of gates in a matter of minutes. 
For example, our basic implementation processed a circuit for \distinct\ 
with close to 16 million gates in under $9$ minutes, or close
to 30,000 gates per second.  
However, since the circuit's size was more than 100 times larger 
than the universe over which the input is drawn, 
this translated to only about 300 items per second. 
The other costs incurred are very low. 
The verifier's
space usage and the communication cost are never more than a few dozen kilobytess, and
the verifier processes close to thirty million updates per second
across all stream lengths. 
The time for $\V$ to run the protocol is negligible compared to the
(already low) time to compute the required low-degree
extension of the input.   

In Section~\ref{app:muggles}, we discuss how adding additional gate
types can reduce the cost of circuit checking. 
We demonstrate experimentally that adding gates which compute the $8$th power ($\mathbin{\char`\^}8$)
or the $16$th power ($\mathbin{\char`\^}16$) of their inputs
achieves substantial reductions in the size of the circuits needed. 
For \distinct, this reduced 
the number of rounds  by nearly a factor of three, the prover
time by close to $20\%$, and the overall communication cost by close to
$30\%$. 
We also discuss in Section~\ref{app:muggles} how to (conceptually)
replace a binary tree of addition gates with a single $\oplus$ gate of
very large fan-in which sums all its inputs. 
For \distinct, this optimization further reduced both communication
and number of rounds by $10$-$20\%$. 
The effect of $\oplus$ gates was much more pronounced for
\sjs, where we saw an order of magnitude reduction in the number of
rounds, and 5-fold reduction in communication cost. 
The change was larger here because the addition gates
represent a much larger fraction of the gates in \sjs\ circuits than 
in \distinct\ circuits.

\begin{figure*}[t]
\newlength{\figwidthdie}
\setlength{\figwidthdie}{0.34\textwidth}
\centering
\subfigure[Verifier's time]{
\includegraphics[width=\figwidthdie]{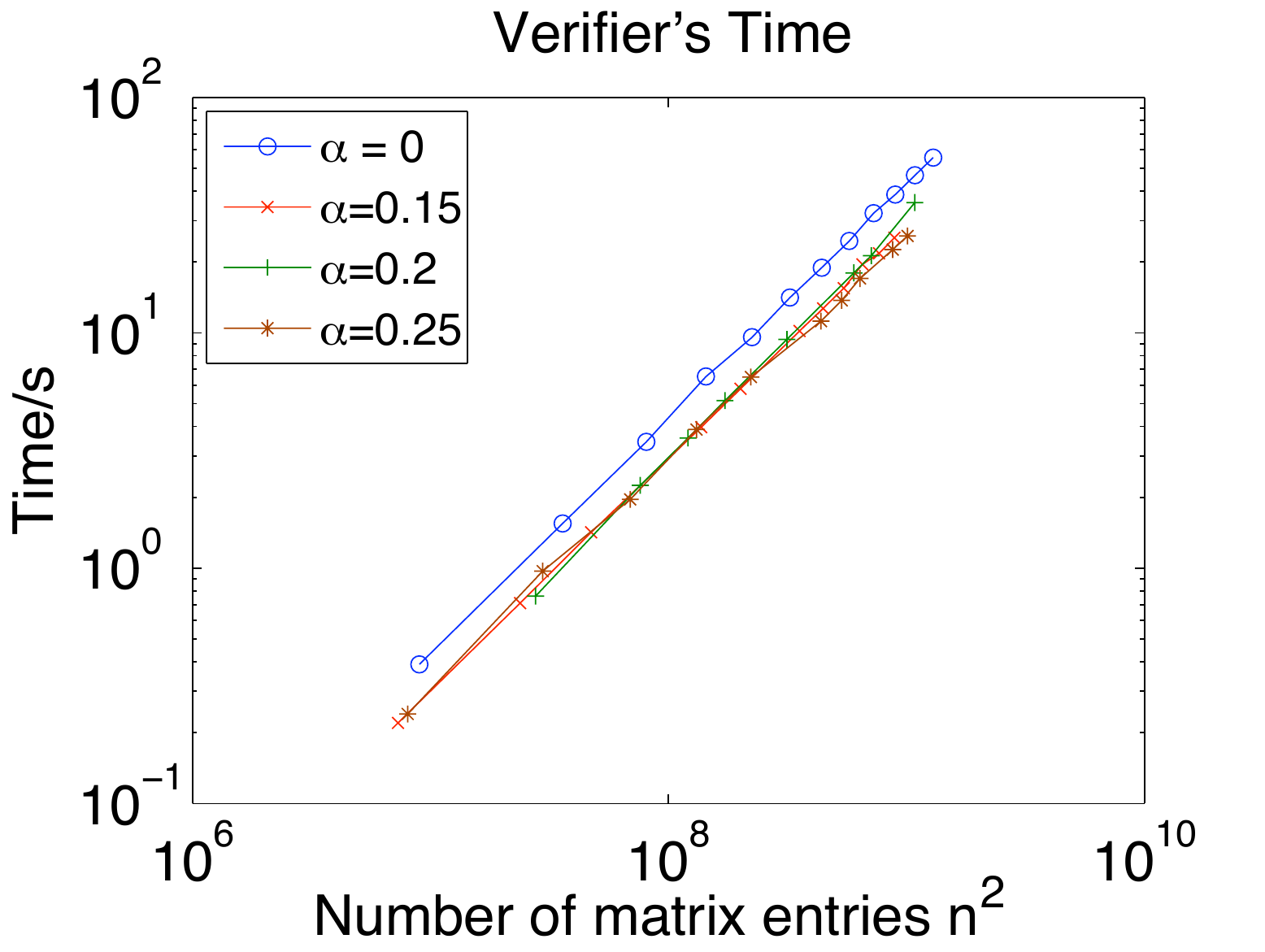}
\label{fig:timevmatvec}
}%
\subfigure[Prover's time]{
\includegraphics[width=\figwidthdie]{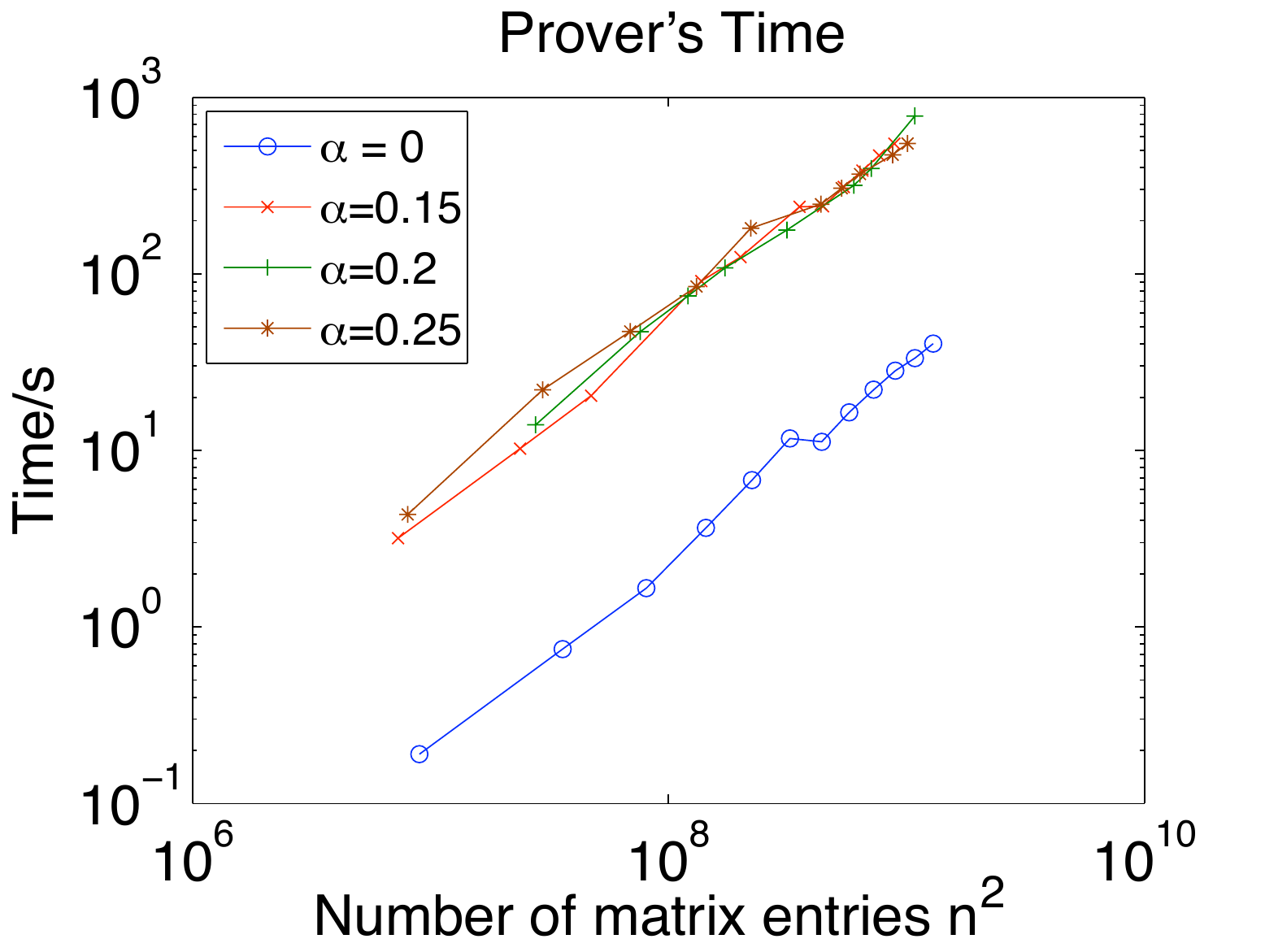}
\label{fig:timepmatvec}
}
\subfigure[Space cost]{
\includegraphics[width=\figwidthdie]{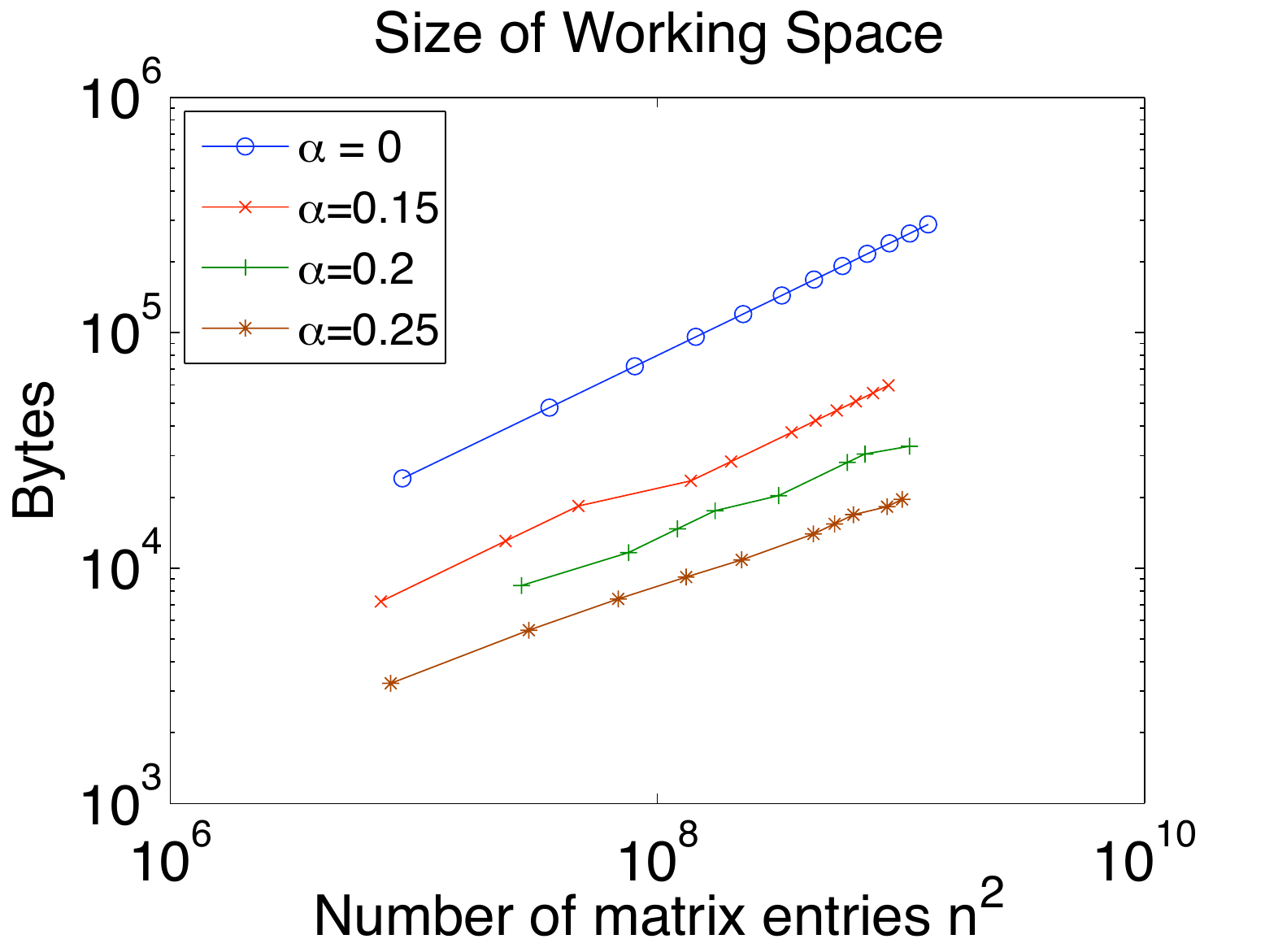}
\label{fig:spacematvec}
}%
\subfigure[Communication cost]{
\includegraphics[width=\figwidthdie]{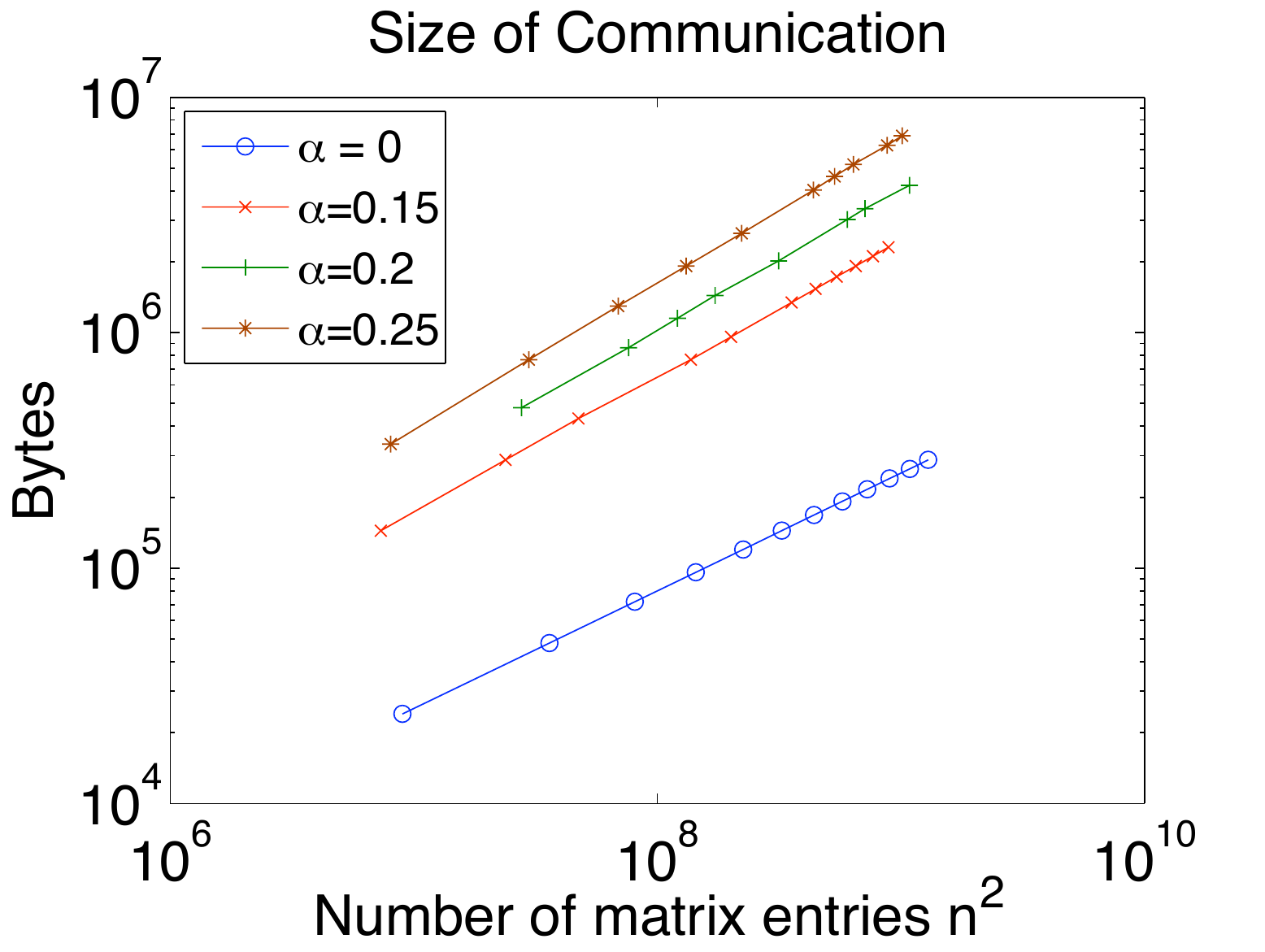}
\label{fig:commatvec}
}
\caption{Experiments on non-interactive \matvect\ protocols.}
\label{fig:matvec}
\end{figure*}

\subsection{Specialized Protocols}
We now describe our experiments with specialized protocols on a
problem-by-problem basis.  
We find that specialized interactive protocols improve over the
general-purpose construction by several orders of magnitude. 
Moreover, we demonstrate that the FFT techniques of Section \ref{sec:FFT} yield
\emph{non-interactive} protocols that easily scale to streams with
billions of updates, improving 
over previous implementations by three orders of magnitude.
The protocols are of various types: 
  the basic multi-round protocols based on sum-check from \cite{pods}
  (MRS); 
  multi-round protocols which use linearization from
  Section~\ref{sec:distinct} (LIN);
  multi-round protocols based on circuit checking described in
  Section~\ref{sec:muggles} (CC); 
  the basic non-interactive protocols from \cite{annotations} (NI);
  and the faster implementation of these protocols via FFT in
  Section~\ref{sec:FFT} (NI-FFT). 

\medskip
\noindent \emph{\sjs}: 
There are four known protocols for \sjs: 
  one obtained via the general-purpose circuit-checking approach (CC),
a specialized interactive protocol due to \cite{pods} (MRS), 
a naive implementation of the non-interactive protocol due to
\cite{annotations} (NI),
and a non-interactive implementation based on our FFT techniques
developed in Section \ref{sec:FFT} (NI-FFT). 
The results for CC are for our optimized implementation using 
$\oplus$ gates.
Figures \ref{fig:timev} and \ref{fig:space} illustrate the verifier's
time and space costs 
for all four protocols, while 
Figure \ref{fig:timep} illustrates the prover's runtime
for these protocols. 
We used implementations of NI and MRS protocols for \sjs\ due to
\cite{pods}.
Note that in the case of NI and NI-FFT, the verifier behaves {\em
  identically}: the prover computes the same messages in both cases,
but more quickly using FFT. 

The main observation from Figures \ref{fig:timev} and
\ref{fig:space} is that the verifier's 
costs are extremely low for all four protocols. 
$\V$ processed over 20 million items/s across all stream lengths for
all protocols. 
The space usage and communication cost for both interactive protocols (CC and MRS) 
is less than 1 kilobyte across all stream lengths tested, while the space usage for
the non-interactive case is much larger but still reasonable
(comfortably under a megabyte even for stream lengths in excess of 1
billion). 

\eat{
Figures \ref{fig:timev}-\ref{fig:space} show the behavior of the
multi-round protocol of \cite{pods} and improved non-interactive
protocol for \sjs\ described in Section \ref{sec:FFT}.
First, Figure \ref{fig:timev} shows the time for the verifier to
process the stream as the domain size increases.  
Both show a linear trend (here, plotted on log scale).  
Moreover, both take roughly the
same time, with the multi-round verifier processing 20-21 million
items per second, and the non-interactive verifier processing 38-40
million.  
The similarity is not surprising: both methods are taking
each element of the stream and computing the product of the frequency
with a function of the element's index $i$ and the random parameter
$r$.  
The effort in computing this function is roughly similar in both
cases.  The non-interactive verifier has a slight advantage, since it
can compute and use lookup tables within the $O(\sqrt{n})$ space
bound, while the multi-round verifier limited to logarithmic space
must recompute some values multiple times.
}
%

Figure \ref{fig:timep} shows a clear separation between the four methods in
$\P$'s effort in generating the proof. 
For large streams, it is clear that NI is not scalable,
with $\P$'s runtime growing like $n^{3/2}$; 
this implementation failed to process streams larger than about 40 million
updates. 
In contrast, the FFT-based implementation of the non-interactive
protocol processed between
$350,000$ and $750,000$ items per second for all tested values of $n$,
even for values of $n$ well into the billions.  
Thus, the FFT techniques of Section \ref{sec:FFT}
speed up $\P$'s  computation by several orders of magnitude compared
to the naive implementation, and allowed the protocol  
to easily scale to streams with billions of items. 
As mentioned in Section \ref{sec:FFT}, a wide variety of more complicated 
protocols use this protocols as a subroutine, and therefore these
non-interactive techniques are as powerful as they are general.  

For the multi-round protocols, 
circuit checking (CC) eventually outpaces NI, and scales linearly: 
the CC prover processed about 20,000 items per second
across all stream lengths. 
Finally, the multi-round prover processed 20-21 million items
per second.
We conclude that special-purpose protocols 
should have substantial value, as our specialized non-interactive protocol
was faster than Circuit Checking by more than an order of magnitude, and
the specialized interactive protocol was faster by two orders of magnitude. 

\eat{
With the exception of the naive non-interactive implementation, the prover's runtime 
for all methods
grows essentially linearly with input size, and even the
non-interactive prover is fast enough to be practical in many
settings.  The cost in the multi-round case is lower than the single
round case by a factor of 20-50 depending on input size.  Across all
values of $n$, the multi-round prover processed 20-21 million items
per second, while the non-interactive prover processed between
$350,000$ and $750,000$ items per second for tested values of $n$.

Figure \ref{fig:F2Second} compares the runtime of our improved prover
described in Section \ref{sec:FFT} to the runtime of the prover as implemented 
in \cite{pods}. We see a clear difference between the two protocols. 
It takes minutes to process inputs with $n=2^{20}$ 
with the implementation of \cite{pods}, whereas the same data requires 
less than two seconds when using our FFT techniques. 
Worse, the earlier prover runtime grows as $n^{3/2}$, so
doubling the input size increases the runtime by a factor of 2.8.
In contrast, the runtime of our implementation grows nearly linearly.
}

\newlength{\figwidththree}
\setlength{\figwidththree}{0.33\textwidth}
\begin{figure*}[t]
\hspace*{-7mm}
\subfigure[Prover's time]{
\includegraphics[width=\figwidththree]{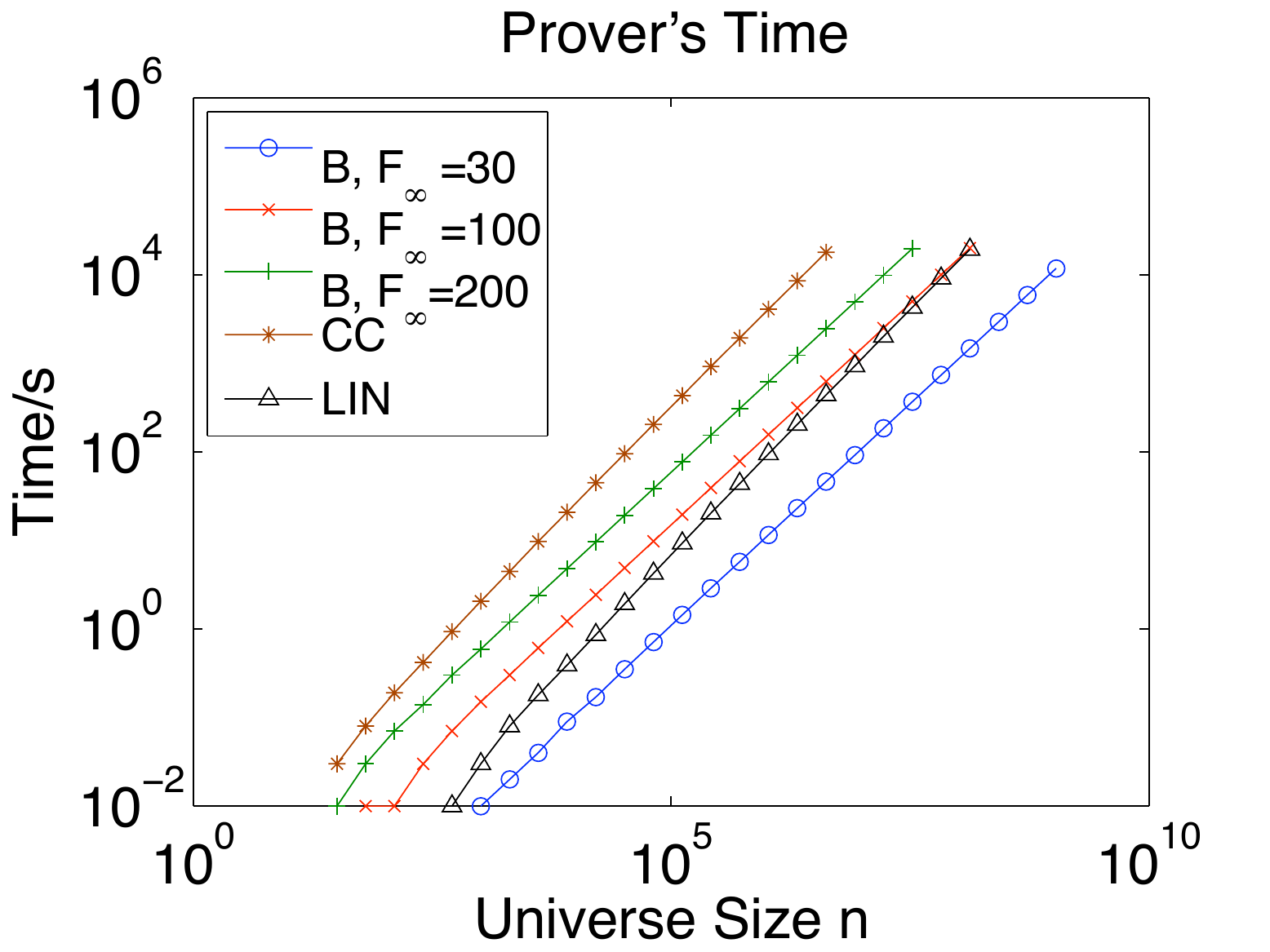}
\label{fig:F0fixedFmax}
}%
\subfigure[Verifier's time]{
\includegraphics[width=\figwidththree]{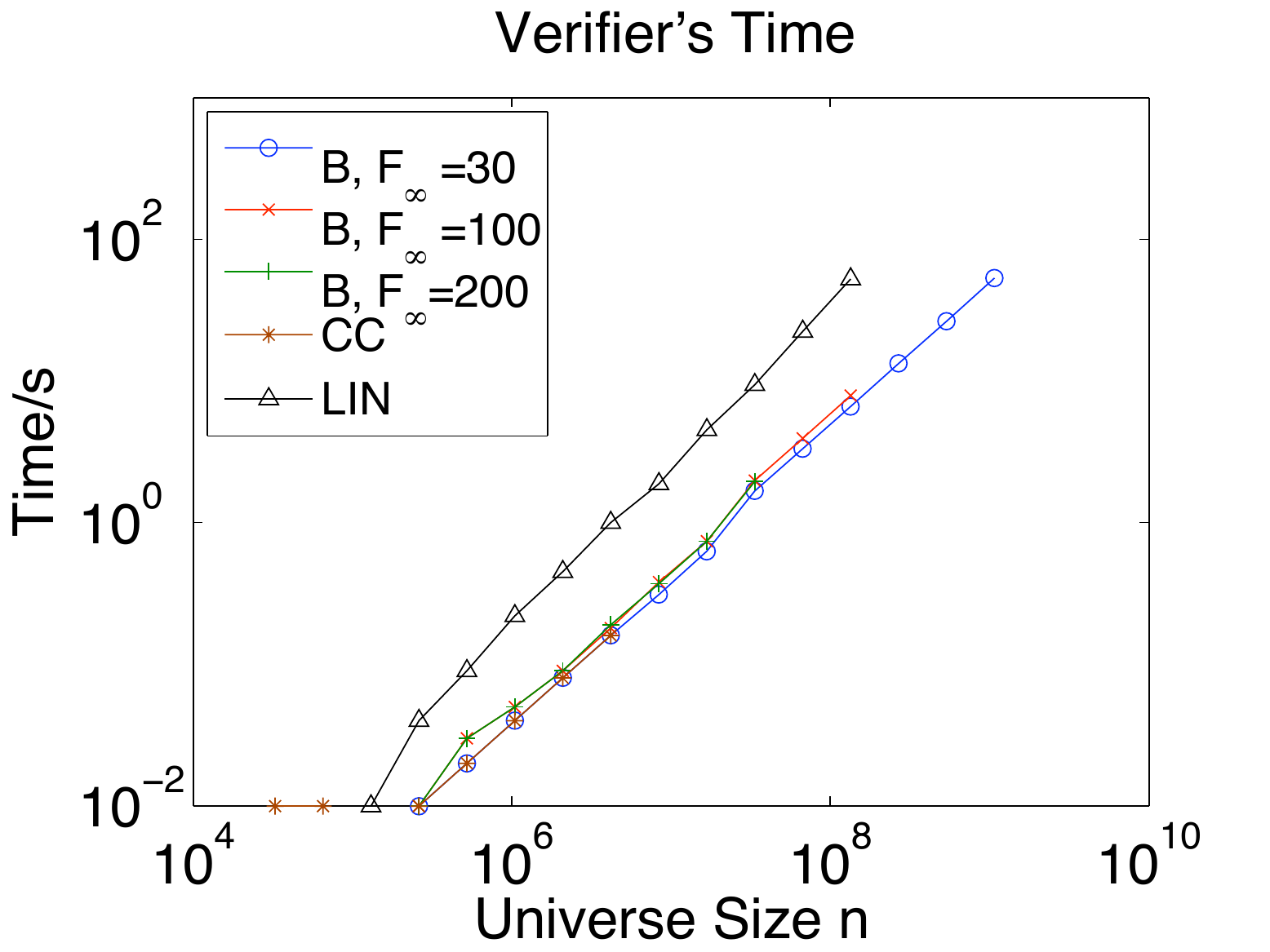}
\label{fig:F0IP}
}%
\subfigure[Communication cost]{
\includegraphics[width=\figwidththree]{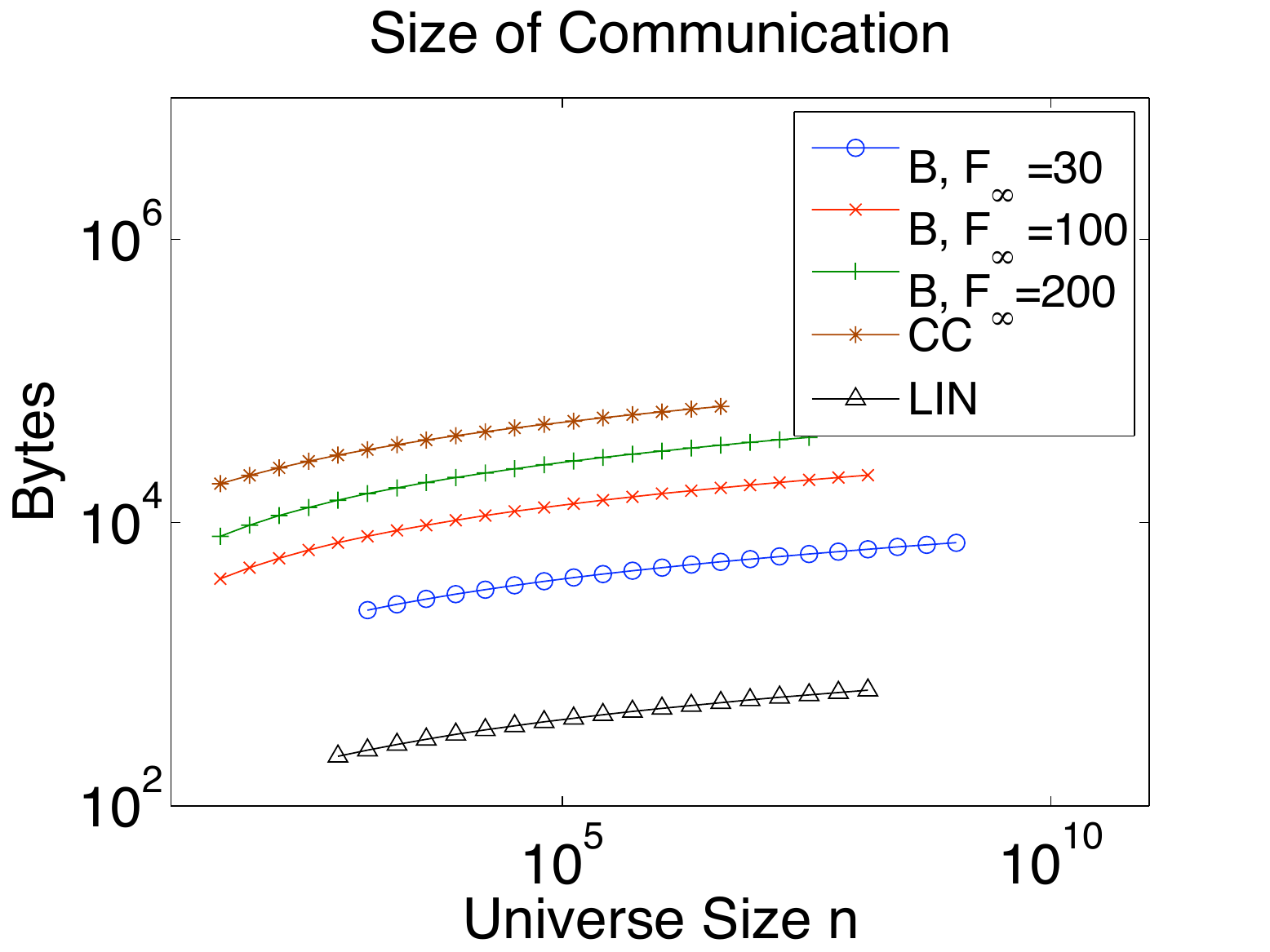}
\label{fig:F0com}
}%
\caption{Experimental results for \distinct.}
\label{fig:svplots}
\end{figure*}
\medskip
\noindent \emph{\matvect:}
Figure \ref{fig:matvec} shows the behavior of our FFT-based
implementation of the $(n^{1+\alpha}, n^{1-\alpha})$ non-interactive protocol for \matvect\ 
described in Section~\ref{sec:FFT}. 
Recall that the parameter $\alpha$ allows us to tradeoff between
communication and space used by the verifier. 
A convenient (and previously unremarked on) feature of this
protocol is that when $\alpha=0$, the honest prover's
 message consists simply of the vector $\b$. 
Consequently, we obtain an $(n, n)$ protocol for which the prover
can handle enormous throughputs: 
$30$-$50$ million items/second as evidenced in Figure
\ref{fig:timepmatvec}.
In outsourcing settings where one can tolerate space usage $O(n)$ for
the verifier, this protocol is truly ideal, as the prover need do
nothing more than solve the problem, and the verifier's computation
consists only of maintaining $n$ fingerprints. 
That is, this $(n, n)$ protocol allows the user to obtain a strong
security guarantee on the integrity of the query, almost for
free.
Note that for this problem, the size of the input is
  $O(n^2)$ for an $n \times n$ matrix, so $O(n)$ space at the verifier
  is still much smaller than the full input size.

The behavior becomes more interesting when we set $\alpha > 0$---in
this case, in addition to providing the correct answer, 
the prover has to do non-trivial computation to prove correctness. 
Because lower values of $\alpha$ mean less space but
more communication (see Figure \ref{fig:spacematvec}),  
setting $\alpha > 0$ may be needed
when the verifier is severely space-limited. 
It may also be necessary when the matrix is very wide:
in full generality the protocol has communication and
space cost $(mn^{\alpha}, n^{1-\alpha})$ for an $m \times n$ matrix. 
We show how different costs vary as a function of $\alpha$: 
$\V$'s time to process the input (Figure \ref{fig:timevmatvec}), 
$\P$'s time (Figure \ref{fig:timepmatvec}),
the communication cost (Figure \ref{fig:commatvec}),
and the space used by $\V$ (Figure \ref{fig:spacematvec}).
Across all values of $\alpha$, 
$\P$ can process in excess of 1 million items per second using our
FFT techniques. 
The verifier runs over the stream slightly faster for higher values
of $\alpha$, because $\V$  maintains fewer fingerprints for larger
$\alpha$'s. 
When $\alpha=0$, $\V$ processed about 20 million
items per second, and when $\alpha=.25$, 
$\V$ processed in excess of 30 million items per second. 
For concreteness, Table \ref{tabmatvec} displays the costs of the protocol when 
run on matrices of size 10,000 $\times$ 10,000. 
\eat{
Translating the space bounds in Table~\ref{tab:nip}, 
for square matrices with 1 billion entries, 
$\alpha=0$  equates to about 250KB in both space and communication, while 
setting $\alpha=0.25$ means $\V$ uses 20KB space, and receives 6.5MB
of messages (processed as a stream).}

\begin{table}[t]
\begin{center}
\begin{tabular}{ | c | c | c | c | c | }
\hline
$\alpha$ & Space &  Comm   & $\P$ time (s) & $\V$ time (s)     \\  
\hline
0 & 78.1 KB & 78.1 KB & ~1.6 & 4.3 \\
.15 & 19.9 KB & 468.8 KB & 33.9 & 3.0 \\
.20 & 12.8 KB & 937.5 KB & 58.9 & 2.8 \\
.25 & 7.8 KB & 1.52 MB & 61.5 & 2.6 \\
\hline

\end{tabular}
\caption{Non-interactive \matvect\ results on matrices of size 10,000 $\times$ 10,000 (763 MBs of data).}
\label{tabmatvec}
\end{center}
\end{table}

\medskip
\noindent \emph{\distinct}:
We implemented the $(\log u, \sqrt{n}\log u)$ interactive protocol of
\cite{pods} described at the start of Section~\ref{sec:distinct}, 
which we refer to as the {\em bounded } protocol (B), since it uses a
bound on \Finf, the maximum frequency of any item. 
We compare this to the new {\em Linearization} based protocol (LIN) from
Section~\ref{sec:linear}, as well as to the circuit checking approach (CC) of Section~\ref{sec:pattern}.
The circuit-checking results shown are from our optimized implementation using $\mathbin{\char`\^}8$ gates.

Our focus is primarily on $\P$'s runtime, since 
we find that the bounded protocol is impractical for
general streams: $\P$'s runtime is $\Theta(n^2)$. 
However, recall from Section \ref{sec:distinct} that 
$\P$'s run time in the bounded protocol can be made $O(\Finf^2 n)$ when 
there is an \emph{a priori} upper bound on \Finf, 
or equivalently when $\V$'s memory is at least $m/\Finf$ for streams
of length $m$.
Figure \ref{fig:F0fixedFmax} shows $\P$'s runtime for
the bounded protocol as a function of the universe size $n$, 
for different bounds on \Finf. 

Figure~\ref{fig:F0fixedFmax} shows
that for fixed \Finf, the prover's runtime in the bounded protocol
grows linearly in $n$ as expected.  
When \Finf\ is very low, the protocol achieves reasonable throughputs,
but as \Finf\ grows the runtime rapidly becomes prohibitive.  
For example, $\Finf=30$ gives about 80,000
items per second, while $\Finf=200$ results in just 1,600
items/second.  
It is clear that this protocol will be unacceptably slow for realistic
streams where \Finf\ is in the thousands or larger.

In contrast, $\P$'s runtime in the linearization and circuit checking
protocols is independent of $F_{\infty}$. 
For linearization,
$\P$'s runtime grows slightly
super-linearly in $n$ (it is $\Theta(n \log^2 n)$ as shown in Section
\ref{sec:distinct}), and as a result the processing speed decreases
slowly as the stream length increases (see Figure \ref{fig:F0IP}). 
For short streams (e.g. $n=2^{16}$), $\P$ handles about 17,000
items/second. 
For $n=2^{24}$, $\P$ handles about 8,000 items per second. 
Extrapolating the behavior to streams of length about 1
billion, $\P$ should handle about 4,500 items/second. 
These results are broadly consistent with its theoretical $\Theta(n\log^2 n)$
running-time bound, 
and represents a substantial improvement over 
the bounded protocol and the circuit checking protocol.
In the circuit checking protocol $\P$ processes only 200-300 items per second 
across all stream lengths.

Note, however, that the overhead for the verifier in all three protocols is very light,
making the costs compelling from $\V$'s perspective. 
In all protocols $\V$'s space was always well under 1KB; this cost was so low for all three protocols that we have
omitted the corresponding plot.
For the circuit checking and bounded protocols, $\V$ processed about 20 million updates per second,
while for the linearization protocol, $\V$ processed 3-5 million items/second.
The verifier in the bounded and circuit-checking protocols   
is faster than in the linearization protocol because, in the first two, $\V$ only requires evaluating a $\log n$-variate
polynomial at a random point,
while the linearization protocol requires evaluating a
 $\log n + \log m$-variate polynomial at a random point. 
The communication requirement  grows larger for circuit
checking and the bounded protocol, with the former approaching 100 KBs
for universes of size 10 million, and the latter approaching similar
amounts of communication when $\Finf=200$. 
In contrast, the communication under linearization was an
order of magnitude lower, never more than a few KBs on all streams tested.

In summary, 
the bounded protocol may be preferable when $\Finf$ is at most a very small
constant (less than about 30); otherwise,  
the linearization protocol dominates, with the only downside being
decreased throughput of the verifier. 

\medskip
\noindent
\pmw:
Our experiments on pattern matching showed broadly the same relative trends as
for \distinct\ and are omitted for brevity. 


\subsection{Parallel Implementations}
The prover's computations in all of the non-interactive
protocols studied here are highly parallelizable, as noted
previously. 
Indeed, using just three OpenMP\footnote{\url{http://www.openmp.org}} statements, we were able to achieve
more than a 7-fold speedup over the sequential implementation 
of the FFT protocol, by using all 8 cores of the multi-core machine
our experiments were run on. Consequently, with 8 processors, the ratio 
between the speed of the MR and NI-FFT
protocols for \sjs\ drops from 20-60 to 3-8. 
%
%
In theory, the interactive $F_2$ protocol is just as easy to
parallelize as the non-interactive protocol; however, we did not find
this to be the case in practice. The prover's computations in the
multi-round protocol are so light-weight (as evidenced by its very
high throughput) that memory access forms the principle
bottleneck.  In our test machine, all cores share a single pipe to
memory, and the bottleneck remains.  In other scenarios, such as each
core having a separate pipeline to memory, multiple cores might yield
more substantial speedups.



\eat{We saw that logarithmic factors make all the difference in practice when
dealing with large data or field size. The difference between the
extremely practical multi-round \sjs\ protocol, which
handle millions of items per second regardless of stream length, and
our new protocol for \distinct\ which handles several thousand
items per second, is a $\log^2 n$ factor (both protocols possess
small hidden constants). Likewise, the principle bottleneck in the
construction of \cite{muggles} is the polylogarithmic blowup in the
number of messages required for each ``non-arithmetic" operation.
\begin{itemize}
\item bullet points if necessary
\end{itemize}
}

{\fussy
\section{Conclusion and Future Directions}}

The ideas and techniques from interactive proof systems have
transformed the landscape of computational complexity over the last
two decades \cite{babai85,gmr85}.  Yet they have had relatively little
practical impact thus far in the area of delegated computation. 
In this paper, we demonstrated that, when combined with significant engineering, interactive proof
systems have sufficiently evolved to yield
protocols suitable for everyday use.

\edit{
A particularly encouraging feature of our experimental results is that $\V$'s runtime
is dominated by the time required to evaluate the LDE of the input at a point $\r$. For the 
low-complexity (linear or near
linear time) computations we experimented on, this cost is actually comparable to the time required to solve the problem
without a prover, assuming $\V$ had enough memory to store the input. But
if we were to
run our implementation on problems requiring superlinear time to
solve, then $\V$ would save significant time as well as space
(compared to solving the problem without a prover). 

Moreover, if the cost of the LDE computation
can be amortized over many queries, then $\V$ will save time as well as space even for very low-complexity functions.
This is indeed possible for our non-interactive protocols, as there is
\emph{no} leakage of information from $\V$ to $\P$ as long as $\P$ does not learn
whether $\V$ accepts or rejects after each query;
soundness is therefore maintained even if $\V$ uses the same $\r$ in all
instances of the protocol. 

Such amortization for interactive protocols may also be possible in 
cases where $\P$ is not considered malicious, such as a
user simply trying to detect a buggy algorithm. In this setting it is reasonable to
use the same location $\r$ in all instances of the protocol even
though soundness is not maintained theoretically. Thus, in these
realistic situations, the amortized time cost to the verifier can be
considerably sublinear in the input length, and our protocols will save the verifier both time and space. 
}

The next step is to further advance the boundary of practicality. 
The chief obstacle for more general systems is the requirement of a
circuit representation for computations, and the superlinear
dependence of the prover's time on the size of the circuit. 
Various approaches offer themselves: either to design protocols which
circumvent this circuit representation, or to improve the throughput
by taking greater advantage of the inherent parallelism in the
prover's work, e.g. via GPU implementation.

\subsection*{Acknowledgments} 
We thank Owen Arden, Varun Kanade, Guy Rothblum, Thomas Steinke, and Salil Vadhan
for helpful discussions. We also thank Shafi Goldwasser, Yael Tauman Kalai, and Guy Rothblum for sharing the
full version of \cite{muggles}.

\newpage
%

\newpage
\appendix
\section{Details for Theorems {\ref{OPT1}-\ref{OPT2}}}
\label{fullapp:muggles}

\graham{In this Appendix, we spell out the details of our
efficient instantiation of the construction of \cite{muggles}. 
Our results ensure that the prover can be implemented efficiently, and that
the verifier can be implemented very efficiently for a large class of circuits.} 

\subsection{Notation and Background} 
We adhere closely to the notation of \cite{muggles}. 
We are given an arithmetic circuit $C$ \graham{of gates with} fan-in 2
over the field 
$\mathbb{F}$. $C$ is in layered form and has size $S(n)$ and depth
$d(n)$, where $n$ is the number of input wires. For presentation
purposes, assume that all layers of the circuit have \graham{at most} $n$
gates, and write $v=\log n$. 
For each $1 \leq i \leq d$, we associate the $j$'th gate at layer $i$ of $C$ with the $v$-bit binary representation of $j$, and for $i > 1$, we define two functions,
$add_i, mult_i : \{0, 1\}^{3v} \rightarrow \{0, 1\}$ which together constitute the wiring predicate of layer $i$ of $C$. Specifically, these functions take as input three gate labels $(j_1, j_2, j_3)$, and return 1 if gate $j_1$ at layer $i-1$ is the addition (respectively, multiplication) of gates $j_2$ and $j_3$ at layer $i$, and return 0 otherwise. We let $\tilde{\text{add}}_i, \tilde{\text{mult}}_i: \mathbb{F}^{3v} \rightarrow \mathbb{F}$ denote the \emph{multilinear extensions} of 
$add_i$ and $mult_i$ respectively. That is, $\tilde{\text{add}}_i$ and $\tilde{\text{mult}}_i$ are the unique multilinear polynomials over $\mathbb{F}$ that agree with $add_i$ and $mult_i$ at all values in $\{0, 1\}^{3v}$.

We also define a function $V_i: \{0, 1\}^v \rightarrow \mathbb{F}$ to represent
the values of the gates at layer $i$. That is, $V_i(j)$ equals the value of gate $j$ at layer $i$.
Let $\tilde{V}_i:\mathbb{F}^v \rightarrow \mathbb{F}$ denote the multilinear extension of $V_i$.

Recall from Section \ref{sec:pattern} that at iteration $i$ of the protocol of \cite{muggles}, the sum-check protocol
is applied to a certain $3v$-variate polynomial $f_i$. We are ready to give the definition of $f_i$ as promised. Given a vector $x \in \mathbb{F}^{3v}$, write $\p=(x_1,\dots, x_v)$, $\omega_1=(x_{v+1},\dots, x_{2v})$ and $\omega_2=(x_{2v+1},\dots, x_{3v})$. Then we define

\begin{align} \label{eq4} f_i(\p, \omega_1, \omega_2) := \beta(p) \left(\tilde{\text{add}}_i(\p, \omega_1, \omega_2) (\tilde{V}_i(\omega_1) + \tilde{V}_i(\omega_2)
)
+  \tilde{\text{mult}}_i(\p, \omega_1, \omega_2)  \tilde{V}_i(\omega_1) \tilde{V}_i(\omega_2)\right).\end{align}

Here, $\tilde{\text{add}}_i, \tilde{\text{mult}}_i,$ and $\tilde{V}_i$ are as above, and $\beta(p)$ is a certain polynomial that depends only on $p$. 

\subsection{Making $\P$ Run in Time $O(S(n) \log S(n))$}

In this subsection we show how to engineer an efficient prover.
First we give an informal outline, then go on to make this more
precise. 

\subsubsection{High-level Outline}
In the $j$'th round of the sum-check protocol applied to $f_i$, 
$\P$ is required to send the \emph{univariate} polynomial
\begin{align*} g_j(X_j) =
& \!\!\!\!\!\!\!\!\!\sum_{(x_{j+1},\dots,\!x_{3v})\in\{0,1\}^{3v-j}} \!\!\!\!
f_i(r^{(i)}_1, \dots, r^{(i)}_{j-1}, X_j, x_{j+1}, \dots, x_{3v}).\end{align*} Theorems \ref{OPT1}, \ref{thm:online}, and \ref{OPT2} rely on the observation that, when $\tilde{\text{add}}_i$ and $\tilde{\text{mult}}_i$ are multilinear extensions, rather than arbitrary low-degree extensions, then each gate at layers $i$ and $i-1$ contributes to exactly one term in the sum. 

More specifically, the key observation is that the multilinear extension of the wiring predicate acts as a sum of \emph{variable-wise} indicator functions on boolean-valued variables, with one indicator function for each gate at the layer of interest. At any round $j$ of the sum-check protocol, the ``unbound'' variables (i.e., those appearing in the sum defining $g_j$) still only range over values in $\{0,1\}$, and thus each gate $\y$ at the current layer of the circuit still contributes to only one term in the sum in intermediate rounds. Namely, $\y$ contributes to the unique term of the sum that agrees with the trailing bits in the binary representation of $\y$, despite the fact that ``bound'' variables may take values outside of $\{0,1\}$. 


\subsubsection{Decomposing $\tilde{\text{add}}_i$ and $\tilde{\text{mult}}_i$ as Sums of Variable-wise Indicator Functions}
Since $\tilde{\text{add}}_i$ and $\tilde{\text{mult}}_i$ are the multilinear extensions of the wiring predicate, we can write them explicitly as follows.

For $\y \in\{0,1\}^{3v}$ let $\chi_\y(x_1, \dots, x_{3v})=\prod_{k=1}^{3v} \chi_{y_k}(x_k)$, where $\chi_{0}(x_k)=1-x_k$ and $\chi_1(x_k) = x_k$. $\chi_\y$ 
is the unique multilinear polynomial that takes $\y \in \{0,1\}^{3v}$ to 1 and all other values in $\{0,1\}^{3v}$ to 0, i.e., it is the multilinear extension of the indicator function for boolean vector $\y$. 

Notice that if $(x_{j+1}, \dots, x_{3v}) \in \{0, 1\}^{3v-j}$, then for any $(r_1, \dots r_j) \in \mathbb{F}^j$, 
\begin{flalign} \label{eq1}
& \chi_y(r_1, \dots, r_j, x_{j+1}, \dots, x_{3v})=
 \begin{cases}
    \prod_{l=1}^j \chi_{y_l}(r_l), & \text{if } x_k=y_k \text{ for all } k\geq j+1 .\\
    0, & \text{otherwise}.
  \end{cases}
\end{flalign}

Informally, Equation \eqref{eq1} implies that one may think of $\chi_\y$ acting as a \emph{variable-wise} indicator function on boolean-valued variables.

Since $\tilde{\text{add}}_i$ and $\tilde{\text{mult}}_i$ are multilinear extensions, they can be written as a sum of these $\chi_y$ functions, 
where each gate $y$ at layer $i-1$ contributes a term $\chi_y$ to the sum. That is,

\begin{equation} \label{eqadd} \tilde{\text{add}}_i(x_1, \dots, x_{3v})=\sum_{\text{add gates } \y \text{ at layer } i-1}
\!\!\!\!\!\!\!\!\chi_y(x_1,\dots, x_{3v})\end{equation}

and 

\begin{equation} \label{eqmult} \tilde{\text{mult}}_i(x_1,\!\dots,\!x_{3v})=\sum_{\text{mult gates }\!\y\!\text{ at layer } i-1}
\!\!\!\!\!\!\!\!\chi_y(x_1, \dots , x_{3v}).\end{equation}

It is straightforward to observe the expressions on the right hand sides of Equations \eqref{eqadd} and \eqref{eqmult} are multilinear polynomials that agree with $\text{add}_i$ and $\text{mult}_i$ on boolean-valued inputs, and hence the right hand sides are equal to \emph{the} multilinear extensions of $\text{add}_i$ and $\text{mult}_i$ respectively.

For any vector $\x=(x_{j+1}, \dots x_{3v}) \in \{0, 1\}^{3v-j}$, and for any $(r_1, \dots r_j) \in \mathbb{F}^j$,  let $\mathbf{x}^*$ denote the vector $$\mathbf{x}^*:=(r_1, \dots, r_j, x_{j+1}, \dots, x_{3v}) \in \mathbb{F}^{3v},$$ and let $S_\x$ denote the set of gates at layer $i-1$ given by
$\{\y \in \{0,1\}^{3v} : y_k=x_k \text{ for all } k \geq j+1\}$. Equations  \ref{eqadd} and \ref{eqmult} imply that

\begin{equation} \label{eq2} \tilde{\text{add}}_i(\x^*)=\sum_{\text{add gates } \y \in S_\x }
\left( \prod_{l=1}^j \chi_{y_l}(r_l)\right),
\end{equation}
 and similarly
 
 \begin{equation} \label{eq3}
 \tilde{\text{mult}}_i(\x^*)=\sum_{\text{mult gates } \y \in S_\x }
 \left(\prod_{l=1}^j \chi_{y_l}(r_l)\right)
 \end{equation}
 
 \subsubsection{Completing the Calculation}
At round $j$ of this sum-check protocol, the prover must compute the message

$$g_j(X_j) = \sum_{x_{j+1} \dots x_{3v} \in \{0, 1\}^{3v-j}}\!\!\!\!\!f_i(r^{(i)}_1, \dots, r^{(i)}_{j-1}, X_j,
x_{j+1} \dots x_{3v}).$$

Since $g_j$ has degree three if we are using multilinear extensions, it suffices for the prover to send $g_j(r_j)$ for $r_j \in \{0, 1, 2\}$, as these evaluations uniquely define $g_j$.

Using Equations \eqref{eq2} and \eqref{eq3}, 
\graham{we can now easily observe that} 
each gate at layer $i-1$ contributes to exactly
one term in the sum. Specifically, 
for any term $\x=(x_{j+1} \dots x_{3v}) \in \{0, 1\}^{3v-j}$ in the sum, let $\mathbf{x}^*$ denote the vector $$\mathbf{x}^*:=(r^{(i)}_1, \dots, r^{(i)}_j, x_{j+1}, \dots, x_{3v}) \in \mathbb{F}^{3v}$$ as before, and let
$\p^* \in \mathbb{F}^v$ be the first $v$ entries of this vector, $\omega_1^* \in \mathbb{F}^v$ the middle $v$ entries, and $\omega_2^* \in \mathbb{F}^v$ the final $v$ entries. Then combining
Equations \eqref{eq2} and \eqref{eq3} with \eqref{eq4}, we see

\begin{align} \label{eqfinal} f_i(\x^*) &= \beta(p^*) \notag \cdot \\ &\!\!\!\!\!\!\!\!\!\Bigg( \Bigg(\sum_{\text{add gates } \y \in S_\x }  \left(\prod_{l=1}^j \chi_{y_l}(r_l)\right)\Bigg) \left(\tilde{V}_i(\omega_1^*) + \tilde{V}_i(\omega_2^*)\right) \notag \\ &\!\!\!\!\!\!\!\!\!\!\!\!\!\!\!\!\!\!+ \Bigg(\sum_{\text{mult gates } \y \in S_\x }  \left(\prod_{l=1}^j \chi_{y_l}(r_l)\right)\Bigg) \cdot \tilde{V}_i(\omega_1^*) \cdot \tilde{V}_i(\omega_2^*)\Bigg).\end{align}
 
Each gate $\y$ at layer $i-1$ is in $S_{\x}$ for exactly one $\x \in \{0, 1\}^{3v-j}$. Namely, $\x$ is the boolean vector 
equal to the last $3v-j$ bits of the binary representation of $\y$. Denote this vector by $\x(\y)$, and similarly let $\x^*(\y)$, $\p^*(\y)$, $\omega_1^*(\y)$ and $\omega_2^*(\y)$ denote the corresponding vectors implied by $\x(\y)$.

Equation \eqref{eqfinal} implies that $\y$ contributes only to the term $\x(\y)$ of the sum defining $g_j(r_j)$ for $r_j \in \{0, 1, 2\}$. That is,
we may write

\begin{align*} g_j(r_j)  = &
\sum_{\text{ add gates } \y \text{ at layer } i-1}\!\!\!\!\!\!\!\!\!\beta(p^*(\y))\!\left(\prod_{l=1}^j\!\chi_{y_l}(r_l)\right)(\tilde{V}_i(\omega_1^*(\y))\!+\!\tilde{V}_i(\omega_2^*(\y)))\\
&\!\!\!\!\!+\!\!\!\!\!\!\sum_{\text{ mult gates } \y \text{ at layer } i-1}\!\!\!\!\!\!\!\!\!\!\!\!\!\!\beta(p^*(\y))\!\left(\prod_{l=1}^j\!\chi_{y_l}(r_l)\right)\!\cdot\!\tilde{V}_i(\omega_1^*(\y))\!\cdot\!\tilde{V}_i(\omega_2^*(\y)).\end{align*}

Thus, the prover can compute $g_j(0)$, $g_j(1)$, and $g_j(2)$ with a single pass over the gates at layer $i-1$. By a similar calculation, all necessary $\tilde{V}_i(\omega_1)$ and $\tilde{V}_i(\omega_2)$ for each message of the prover can be computed with a single pass over the gates at layer $i$.  
In conclusion, as long as we use the multilinear extension of the circuit's wiring predicate, the prover can compute each message at layer $i$ with a single pass over the gates at layer $i-1$ and a single pass over the gates at layer $i$, performing a constant number of field operations for each gate. Thus, the prover runs in time $O(S(n) \log S(n))$ in total, where $S(n)$ is the size of the circuit.

\subsection{Finishing the Proofs of Theorems \ref{OPT1} and \ref{thm:online}}
We have demonstrated that if the protocol of \cite{muggles} is instantiated with the 
multilinear extensions of the circuits wiring predicate and gate value function, then $\P$
can be made to run in time $O(S(n) \log S(n))$. All that remains in proving Theorem \ref{OPT1} is to show that for any log-space uniform circuit, the verifier can evaluate $\tilde{\text{add}}_i(p, \omega_1, \omega_2)$ and $\tilde{\text{mult}}_i(p, \omega_1, \omega_2)$ in space $O(\log S(n))$. 
This holds because $\V$ can make an ``implicit'' pass over each layer of the
circuit and compute the contribution of each gate to $\tilde{\text{add}}_i$ and $\tilde{\text{mult}}_i$.  
That is, $\V$ considers each gate $y$ in turn, and computes $y$'s contribution to
$\tilde{\text{add}}_i$ and $\tilde{\text{mult}}_i$ using Equations
\eqref{eqadd} and \eqref{eqmult}.
This requires $O(S(n))$ time in total, but only $O(\log S(n))$ space,
since $\V$ never needs to store  an explicit representation of the
circuit. Theorem \ref{OPT1} follows. 

Theorem \ref{thm:online} follows from the additional observation that $\tilde{\text{add}}_i$ and $\tilde{\text{mult}}_i$ do not depend on the input, nor do the random coins of the verifier, and these coins 
uniquely determine the points at which $\V$ must evaluate $\tilde{\text{add}}_i$ and $\tilde{\text{mult}}_i$. Thus, $\V$ can toss all her coins in the pre-processing phase and compute the necessary evaluations of $\tilde{\text{add}}_i$ and $\tilde{\text{mult}}_i$. $\V$ stores the answers and the random coins for use in the online phase. In the online phase, $\V$ only needs to spend $O(1)$ time per round of the protocol to check $\P$'s messages for consistency, and thus $\V$ takes time $O(d(n) \log S(n))$ in the online phase.

In streaming contexts, where $\V$ is more space-constrained than 
time-constrained, this may be acceptable.
However, the solutions we adopt in our experimental implementation correspond
to \graham{the stronger} Theorem \ref{OPT2},
\graham{which further reduces the space and time costs for the verifier.}

\subsection{Discussion and Formal Statement of Theorem \ref{OPT2}}
Now that we have defined the polynomial $f_i$ to which the sum-check protocol is applied
in the $i$'th iteration of the construction of \cite{muggles}, we are ready to state
Theorem \ref{OPT2} formally. 

\begin{theorem}[Formal statement of Theorem \ref{OPT2}.]
 \label{thm:formal} 
Let $C$ be a log-space uniform circuit of size $S(n)$ and depth $d(n)$, and assume there exists an $O(\log(S(n)))$-space,\\
 $\poly\log(S(n))$-time algorithm for evaluating $\tilde{\text{add}}_i$ and $\tilde{\text{mult}}_i$ at a point, for all layers $i$ of the circuit.
Then $\P$ requires $O(S(n) \log S(n))$ time 
to implement the protocol of Theorem~\ref{thm:muggles}
over the entire execution.
$\V$ requires space $O(\log S(n))$ and time 
$O(n \log n + d(n)\poly(\log S(n)))$, where the $O(n \log n)$ term is due to the time required
to evaluate the low-degree extension
of the input at a point. \end{theorem}

The remainder of this section is devoted to discussing the applicability of Theorem \ref{thm:formal}. We believe the assumption that the multilinear polynomials $\tilde{\text{add}}_i$ and $\tilde{\text{mult}}_i$ can be evaluated quickly by a small-space algorithm is mild, in both theory and practice. 
We demonstrate this in three ways. First, we show that all four motivating problems in this
work possess succinct circuits to which Theorem \ref{thm:formal} applies. Second, we identify
a host of other important circuits from the algorithmic literature to
which Theorem \ref{thm:formal} also applies. Third, we apply
Theorem \ref{thm:formal} to a \jedit{complicated} circuit appearing in the proof of 
\cite[Corollary 1]{muggles}, to obtain improved protocols for any language decidable by a (non-deterministic) Turing Machine in small space.

In essence, Theorem \ref{thm:formal} applies to any circuit with a
``highly regular'' wiring pattern; this explains why it applies to
such a wide array of circuits. The details in the remainder of the
section grow lengthy at times, but the thesis is clear:
Theorem \ref{thm:formal} applies to 
most circuits that arise in both practical applications and
theoretical constructions. 

\subsubsection{Wiring Predicates for \sjs, \distinct, \pmw, and \matvect}
\label{sec:4probs}

\newlength{\figwidthapp}
\setlength{\figwidthapp}{0.4\textwidth}
\begin{figure}[t]
\centering
\includegraphics[width=\figwidthapp]{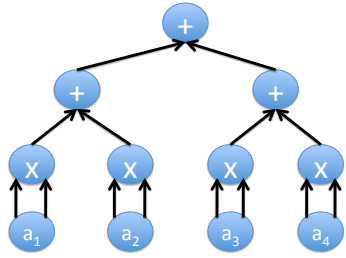}
\caption{A circuit for \sjs\ on 4 inputs.}
\label{fig:F2circ}
\end{figure}

We demonstrate that Theorem \ref{thm:formal} applies to all four circuits described in Section \ref{sec:muggles}.

\begin{enumerate}
\item \textbf{\sjs:}  
Recall that the circuit for \sjs\ had a layer of multiplication gates used for computing the square of each input, and then subsequent levels formed a binary tree of addition gates used to sum up the results. A visual depiction of this circuit on $n=4$ inputs is provided in Figure \ref{fig:F2circ}.

First, consider layer $d-1$ immediately above the input
gates, which consists of multiplication gates used to square each input;  
both the in-neighbors of gate $i$ at layer $d-1$ are equal to
the $i$'th input gate. 
Therefore, if $p=(p_1, \dots, p_v) \in \{0, 1\}^v$ denotes the boolean
representation of a gate at layer $d-1$, 
and $\omega_1=(\omega_{1,1},\!\dots,\!\omega_{1,v})\!\in\!\{0,1\}^v$ and $\omega_2=(\omega_{2,1},\!\dots,\!\omega_{2,v})\!\in\!\{0,1\}^v$
denote the boolean representation of two gates at the input layer, then 
$mult_d$ evaluates to true if and only
if  $p=\omega_1=\omega_2$, while $add_d$ is identically zero.
It is easily seen that the multilinear extension of $mult_d$ is the polynomial
\[\textstyle \tilde{\text{mult}_d}(p,\!\omega_1,\!\omega_2)\!=\!\prod_{j=1}^v\!\big(p_j\omega_{1,j}\omega_{2,j} + (1-p_j)(1-\omega_{1j})(1-\omega_{2,j}) \big),\]
while the multilinear extension of $add_d$ is the zero polynomial.
Clearly, $\tilde{\text{mult}_d}$ can be evaluated at any point in $\mathbb{F}_p^{3v}$
in time and space $O(v) = O(\log n)$.
  
The rest of the circuit for \sjs\ consists of a binary tree of addition gates, which is used to sum up the squared item frequencies. Thus, $\tilde{\text{mult}}_i$ is the zero polynomial for all $i<d$. Meanwhile, for $i<d$ the predicate $add_i(p_1, \omega_1, \omega_2)$ evaluates to 1 if $\omega_1=2p$ and $\omega_2=2p+1$, where here we are interpreting $p$, $\omega_1$, and $\omega_2$ as integers. Thus, it can be seen that
 
 \begin{align*} & \tilde{\text{add}}_i(p, \omega_1, \omega_2) = (1-\omega_{1,1}) \omega_{2,1} \cdot 
 \prod_{j=2}^{v} \left(p_{j-1}\omega_{1,j} \omega_{2,j} + (1-p_{j-1})(1-\omega_{1,j}) (1-\omega_{2,j})\right).\end{align*}
 
Conceptually, the leading factor $(1-\omega_{1,1})\omega_{2,1}$ ensures that $\omega_1$ is even (i.e. its first bit is 0) and $\omega_2$ is odd (i.e. its first bit is 1), while the expression 
\begin{align*} \prod_{j=2}^{n} \left(p_{j-1}\omega_{1,j} \omega_{2,j} + (1-p_{j-1})(1-\omega_{1,j}) (1-\omega_{2,j})\right)\end{align*} 
ensures that the high-order $n-1$ bits of $\omega_1$ and $\omega_2$ agree with the bits of $p$. $\tilde{\text{add}}_i$ is therefore the unique multilinear polynomial evaluating to 1 on boolean inputs $(p, \omega_1, \omega_2)$ if $\omega_1=2p$ and $\omega_2=2p+1$, and evaluating to 0 otherwise.
Clearly $\tilde{\text{add}}_i$ can be evaluated at any point in time and space $O(v)= O(\log n)$.
This completes the description of $\tilde{\text{add}}_i$ and $\tilde{\text{mult}}_i$ for all layers of the circuit for \sjs.

\begin{figure}
\centering
\includegraphics[width=\figwidthapp]{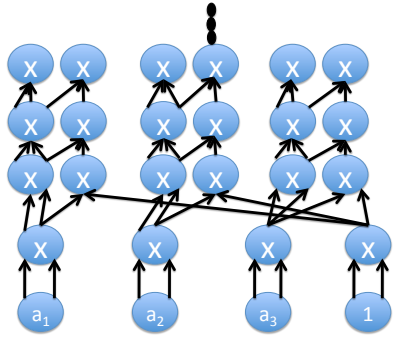}
\caption{The first several layers of a circuit for \distinct\ on three inputs (in place of a fourth input is a ``constant'' gate with value one) over the field $\mathbb{F}_p$ with $p=2^{61}-1$.
The first layer from the bottom computes $a_i^2$ for all $i$. The second layer from the bottom computes $a_i^4$ and $a_i^2$ for all $i$.
The third layer computes $a_i^8$ and $a_i^6 = a_i^4 \times a_i^2$ for all $i$, while the fourth layer computes $a_i^{16}$ and $a_i^{14} = a_i^8 \times a_i^6$ for all $i$. The remaining layers (not shown) have structure identical to the third and fourth layers until the value $a_i^{p-1}$ is computed for all $i$, and the circuit culminates in a binary tree of addition gates.}
\label{fig:F0circ}
\end{figure}

\item \textbf{\distinct:} Recall that for each of the $n$ inputs $a_i$, the circuit for \distinct\ from Section \ref{sec:muggles} computes $a_i^{p-1}$ via $O(\log p)$ multiplications,
and then sums the results via a binary tree of addition gates. We have already seen the wiring predicate for binary trees, so here we only sketch the wiring predicate for the $a_i^{p-1}$ computation, omitting some details for brevity. We do so for the special case of $p=2^{61}-1$, which is the value of $p$ used in
our experiments, as this happens to have a particularly ``regular''
circuit for computing $a^{p-1}$; the calculation would be similar but
\graham{less symmetric} for other values of $p$. 

We may write $p-1=2^{61}-2$, whose binary representation is 60 1s followed by a 0. Thus, $a^{p-1} = \prod_{j=1}^{60} a^{2^j}$. The circuit computing $a^{p-1}$ repeatedly squares $a$, and multiplies together the results ``as it goes''. In more detail, for $i>1$ there are two multiplication gates at each layer $d-i$ of the circuit for computing $a^{p-1}$; the first computes $a^{2^i}$ by squaring the corresponding gate at layer $i-1$, and the second computes $\prod_{j=1}^{i-1} a^{2^{i-1}}$. See Figure \ref{fig:F0circ} for a visual depiction of the first few layers of the \distinct\ circuit.

At a high level then, the wiring predicate $mult_i(p, \omega_1, \omega_2)$ tests equality of $\omega_1$ and $\omega_2$ with two strings that depend on the parity of $p$, as even values of $p$ correspond to gates computing $a^{2^i}$ while odd values correspond to gates computing $\prod_{j=1}^{i-1} a^{2^{i-1}}$. Thus, we may write 

$$\tilde{\text{mult}}_i(p, \omega_1, \omega_2)\!=\!(1-p_1) \chi_{\text{even}}(p, \omega_1, \omega_2)\!+\! p_1\chi_{\text{odd}}(p,\omega_1, \omega_2),$$ 
where $\chi_{\text{odd}}$ and $\chi_{\text{even}}$ are multilinear extensions of the appropriate equality predicates, which do not depend on $p_1$ (we omit a precise definition of $\chi_{\text{odd}}$ and $\chi_{\text{even}}$ for brevity). This can clearly be evaluated in $O(v)$ time and space. 

\item \textbf{ \pmw:} The circuit for \pmw\ is similar to that for \distinct\, so we omit the details for this circuit.

\item \textbf{\matvect:} The circuit described in Section \ref{sec:muggles} for \matvect\ computes $(A\x - \b)_i$ for all $1\leq i\leq n$, and then applies the circuit for \distinct\ to the result. We have already sketched the wiring predicate for \distinct, so we need only describe the wiring predicate of the circuit $C$ computing $(A\x - \b)_i$ for all $1\leq i\leq n$. For presentation purposes, we only describe the wiring predicate for a circuit $C'$ which computes $(A\x)_i$ for all $i$. The wiring predicate for $C'$ is simpler than that of $C$, since $C$ requires some extra gates to ``propagate'' the entries of $\b$ up to the final layer of the circuit, where they are finally used to compute $(A\x - \b)_i$ for all $1\leq i\leq n$. We emphasize that Theorem \ref{thm:formal} applies to the circuit $C$ as well.

Assume $n$ is a power of 2. To simplify the wiring predicate of $C'$, we will treat $C'$ as having $2n^2$ inputs, where  the first $n^2$ inputs of $C'$ are the entries of $A$ in row-major order; and the last $n$ inputs are the entries of the vector $\x$, with all the remaining inputs (between $n^2$ and $2n^2-n$) set to 0 and ignored in subsequent layers. We emphasize that this convention does not increase the costs to either $\P$ or $\V$ in the protocol applied to $C'$.

Each of the $2n^2$ inputs can be specified with $1+2\log n$
bits. Conceptually, the first bit indicates whether the input
specifies an entry of $A$ (a zero indicates yes). The next $\log n$
bits specify the row of $A$, and are zero for any entry of $x$. The
last $n$ bits specify the column of $A$ or the entry of $x$. We
therefore write an input \graham{to a} gate as $h \circ i \circ j$,
where $h \in \{0, 1\}$, $i, j \in \{0, 1\}^n$, and $\circ$ denote
concatenation. 

Layer $d-1$ of $C'$ computes $A_{ij} x_j$ for all $1 \leq i, j \leq n$; 
there are therefore $n^2$ gates at this layer, so each gate can be specified with $2 \log n$ bits. 
This layer consists only of multiplication gates, where the first input to gate $p = i \circ j$ has bit representation $0 \circ i \circ j$, while the second input has bit representation $1 \circ \mathbf{0} \circ j$.
Thus, for $p \in \{0, 1\}^{2 \log n}$, $\omega_1, \omega_2 \in \{0, 1\}^{2 \log n + 1},$ 

$$\tilde{\text{add}}_{d}(p, \omega_1, \omega_2) = 0,$$ while

\begin{align*} &\tilde{\text{mult}}_{d}(p, \omega_1, \omega_2) = \left(1-\omega_{11}\right) \omega_{21} \cdot\\ 
&\!\!\! 
\left(\prod_{k=1}^{\log n}\!\!\left(p_k \omega_{1, k+1}\!+\!\left(1\!-\!p_k\right) \left(1\!-\!\omega_{1, k+1} \right) \right) \left(1\!-\!\omega_{2,k+1}\right)\!\right)\cdot\\
&\!\!\!\!\!\left(\prod_{j=\log n + 1}^{2\log n}\!\!\!\!\left(p_k \omega_{1, k+1} \omega_{2, k+1}\!+\!(1\!-\!p_k) \left(1\!-\!\omega_{1, k+1}\right) \left(1\!-\!\omega_{2, k+1}\right)\right)\! 
\right)\end{align*}

Conceptually, the term $\left(1-\omega_{1,1}\right) \omega_{2,1}$ ensures that the first bit of $\omega_1$ is 0, and the first bit of $\omega_2$  is 1. For $p=i\circ j$, the term $$\prod_{k=1}^{\log n}\!\left(p_k \omega_{1, k+1} + \left(1-p_k\right) \left(1- \omega_{1, k+1} \right) \right) \left(1-\omega_{2,k+1}\right)$$ ensures that the next $\log n$ bits of $\omega_1$ equal $i$, while the corresponding bits of $\omega_2$ are all 0. Finally, the term 
$$\prod_{j=\log n + 1}^{2\log n}\!\!\!\!\!\!\left(p_k \omega_{1, k+1} \omega_{2, k+1}\!+\!\left(1-p_k\right) \left(1\!-\!\omega_{1, k+1}\right) \left(1- \omega_{2, k+1}\right) \right)$$ ensures that the last $\log n$ bits of both $\omega_1$ and $\omega_2$ equal $j$.


Subsequent layers of $C'$ compute $\sum_{j=1}^n A_{ij}x_j$ for each
$1 \leq i \leq n$, which is performed via a binary tree of addition
gates for each $i$.  We have already described the predicate for this
wiring pattern in the paragraph on \sjs. 

\end{enumerate}

\subsubsection{Other Circuits}
Theorem \ref{thm:formal} applies to many other circuits that arise in the algorithms literature. 
Here we provide an incomplete list, sketching the necessary observations for each. 
\begin{enumerate}
\item {\em Matrix Multiplication}. Theorem \ref{thm:formal} applies to the
naive circuit of size $O(n^3)$ and depth $O(\log n)$ for multiplying
two $n \times n$ matrices, which is similar to the circuit $C'$
described in Section \ref{sec:4probs} for \matvect. 
More generally, other multiplication algorithms, such as Strassen's
algorithm, are also amenable to encoding as circuits, reducing the
size to $O(n^{2.807})$ in this case. 
We omit the details of these circuits for brevity. 

\item {\em Rational permutations}. Rational permutations have arisen in the
study of memory hierarchies \cite{ratperm1, ratperm2}, and capture
commonly-used operations such as matrix transposition and
bit-reversal. Formally, a permutation $\Pi$ on $[2^n]$ is rational if
it can be expressed as a permutation $\pi$ on bit positions
i.e. $\Pi((x_1, \dots, x_n)) = (x_{\pi(1)}, \dots,
x_{\pi(n)})$ \cite{ratperm2}.  
There is a two-layer circuit $C$ of size $n$ for \graham{performing any
rational permutation (i.e. producing output wires that are the
permutation of the input wires).}
Let the $0$'th input gate of $C$ be a ``constant gate'' hard-coded to value zero. Each gate $p$ at the non-input layer of $C$ is an addition gate, whose first input is the constant gate, and whose second input is $\Pi(p)$. Then $\tilde{\text{mult}}_1$ is the zero polynomial, while
\begin{align*}  \tilde{\text{add}}_1(p, \omega_1, \omega_2) =  \prod_{j=1}^{\log n} \left(1-\omega_{1j}\right) \left(p_j \omega_{2, \pi(j)} + (1-p_j)(1-\omega_{2, \pi(j)})\right).\end{align*}

Conceptually, the $ \left(1-\omega_{1i}\right)$ term ensures that $\omega_1=0$ while the term $(p_j \omega_{2, \pi(j)} + (1-p_j)(1-\omega_{2, \pi(j)}))$ ensures that $\omega_2 = \Pi(p)$. Clearly, $\tilde{\text{add}}_1$ can be evaluated at a point in $\polylog(n)$ time as long as $\pi(i)$ can be evaluated in $\polylog(n)$ time for $i \in \{0,1\}^{\log \log n}$.

If a rational permutation is used as an intermediate step in a computation represented by a circuit $C'$, then we need not explicitly materialize the above ``rational permutation'' circuit $C$ as an intermediate layer $i$ of the larger circuit $C'$. Rather, we can simply modify the wiring predicate of layer $i$ of $C'$ to directly apply the rational permutation to its variables. That is,
we replace $\tilde{\text{add}}_i(p, \omega_1, \omega_2)$ and  $\tilde{\text{mult}}_i(p, \omega_1, \omega_2)$ with the polynomials 
 $\tilde{\text{add}}_i(p, \Pi(\omega_1), \Pi(\omega_2))$ and  $\tilde{\text{mult}}_i(p, \Pi(\omega_1), \Pi(\omega_2))$. It is easy to see that $\tilde{\text{add}}_i(p, \Pi(\omega_1), \Pi(\omega_2))$ and $\tilde{\text{mult}}_i(p, \Pi(\omega_1), \Pi(\omega_2))$ are multilinear polynomials as long as $\Pi$ is a rational permutation, and these polynomials can be evaluated in $\polylog(n)$ time as long as $\pi(i)$ can be evaluated in $\polylog(n)$ time for $i \in \{0,1\}^{\log \log n}$.

\item {\em Fourier Transform}. Theorem \ref{thm:formal} applies to an arithmetic circuit over the complex field $\mathbb{C}$ computing the standard radix-two decimation-in-time FFT (the most common form of the Cooley-Tukey algorithm \cite{tukey}). 
Let $x \in \mathbb{C}^n$ be the input vector, where $n$ is a power of 2, and let $X \in \mathbb{C}^n$ denote the output vector. The radix-two decimation-in-time FFT relies on the following recursion: Denoting the even-indexed inputs $x_{2k}$ by $E_k$ and the odd-indexed inputs $x_{2k+1}$ by $O_k$, it holds that 
$$X_k = \begin{cases}
E_k + e^{-2\pi ki/n} O_k & \text{if } k \leq n/2\\
E_{k-n/2} + e^{-2\pi ki/n} O_{k-n/2} & \text{if} k > n/2
\end{cases}
$$

\graham{The algorithm is sufficiently well-known that good introductions are
readily available, along with illustrations of a circuit implementing
the above recursion \cite{wiki}.}
Essentially, the circuit performs a bit-reversal on its inputs (which
can be implemented as a rational permutation described above), and
then executes $\log n$ ``stages'', where the $k$'th output of stage
$i$ equals  
\begin{align} \label{eq:twiddle} & V_i(k_1, \dots, k_n) = V_{i-1}(k_1, k_{i-1}, 0, k_i, \dots, k_n) + e^{-2\pi ki/n} V_{i-1}(k_1, \dots, k_{i-1}, 1, k_{i+1}, \dots k_n). \end{align}
Here $V_{i-1}(k)$ denotes the value of the $k$'th output of the previous stage.

The $i$'th stage can thus be implemented with two layers of gates; the first consists only of multiplication gates, and serves to multiply the outputs
of the previous stage by the appropriate twiddle factors (the terms of the form $e^{-2\pi ki/n}$).
The second layer consists only of addition gates, and combines outputs as in Equation \eqref{eq:twiddle}. The wiring predicate of both layers essentially tests whether the $k$'th bit of gate $p$ is 0 or 1, and performs an appropriate equality test depending on the result. We have seen how to 
write equality tests of this form as succinct multilinear polynomials in the paragraph describing the circuit for \distinct\ in Section \ref{sec:4probs}.
\end{enumerate}

\subsubsection{More Efficient Protocols for Space-Bounded Computation}
Our final result of this section is to obtain more efficient protocols for any language decided by a non-deterministic Turing Machine in small space. In the full version of \cite{muggles}, Goldwasser, Kalai, and Rothblum obtain the following result.

\begin{lemma} \label{thm:nl} (\cite{muggles}, full version) Let
$\mathcal{L}$ be any language solvable by a non-deterministic Turing
Machine $T$ in space $s(n) = \Omega(\log n)$ and time $t(n)$. Then
there is an arithmetic circuit $C$ over an extension field of
$\mathbb{F}_2$ computing $L$, where $C$ has size \jedit{$S(n)=\poly(2^{s(n)})$}, and
depth $d(n)=O(s(n) \log t(n))$. Moreover, for $1 \leq i \leq d(n)$,
there exist polynomial extensions $\tilde{\text{add}}_i$ and
$\tilde{\text{mult}}_i$ of the functions $add_i$ and $mult_i$, where
$\tilde{\text{add}}_i$ and $\tilde{\text{mult}}_i$ have degree
$\poly(s(n))$ and can be evaluated at a point using space $O(\log
S(n))$ and time $\poly(s(n))$. 
\end{lemma} 

We show that in fact the circuit $C$ satisfies the following stronger
property: 

\begin{corollary}
\jedit{Let $C$, $add_i$ and $mult_i$ be as in Lemma \ref{thm:nl}.} 
For $1 \leq i < d(n)$, the \emph{multilinear} extensions
 $\tilde{\text{add}}_i$ and $\tilde{\text{mult}}_i$ of the functions
 $add_i$ and $mult_i$, can be evaluated at a point using $O(\log
 S(n))$ words of memory and time $\poly(s(n))$, while
 $\tilde{\text{add}}_{d(n)}$ and $\tilde{\text{mult}}_{d(n)}$ can be
 evaluated at a point using $O(\log S(n))$ words of memory and time
 $O(n \cdot s(n) \log n)$.
\end{corollary}

 Thus, Theorem \ref{thm:formal} implies that in applying the protocol of \cite{muggles} to $C$, the prover can be made to run in time $O(S(n) \log S(n))$, rather than $\poly(S(n))$, with a verifier who uses $O(\log S(n))$ space and runs in time $O(n \cdot s(n) \log n + d(n) \polylog(S(n)))$, where $S(n)$ is the size of $C$. Notice in particular that for any language in $\mathcal{NL}$, the verifier runs in time $O(n \log^2 n)$.

In essence, there are two sources of overhead in the protocol implied by Lemma \ref{thm:nl}, where by overhead we mean the extra computation $\P$ must do to solve the problem \emph{verifiably}, rather than just solving the problem in an an unverifiable manner. First, there is overhead in representing a uniform computation as a (potentially large) circuit $C$ rather than as a non-deterministic Turing Machine $T$. Second, there is additional overhead caused by the fact that in Lemma \ref{thm:nl}, the prover takes time superlinear in the size of $C$. Our results in this section remove the latter source of overhead, or at least reduce it to a logarithmic factor rather than polynomial factor, while maintaining a super-efficient verifier. 

\paragraph{Description of $C$}
In order to present our result, we must first summarize the circuit
$C$ as defined in \cite{muggles}, which can be described as follows.
The non-deterministic Turing Machine $T$ is assumed without loss of generality to have a unique accepting configuration. The circuit $C$ consists of two stages: the first stage computes the adjacency matrix of the configuration graph of $T$ on input $x$, which requires just a single layer of gates, while the second stage determines whether there is a path from the starting configuration of $T$ on input $x$ to the accepting configuration. The second stage determines whether such a path exists by a process resembling repeated squaring of the adjacency matrix of $T$. 

More specifically, closely following the notation in the full version
of \cite{muggles}, a configuration of $T$ can be specified as
a \graham{tuple} $u=(q, i, j, t) \in \{0, 1\}^{g(n)}$, where
$g(n)=O(1) + \log n + \log s(n) + s(n) = O(s(n))$. 
\graham{In this tuple, }
$q$ is a boolean vector describing the machine's state $(O(1)$ bits), $i$ is the boolean representation of the location of the input-tape head ($\log n$ bits), $j$ is the location of the work-tape head ($\log s(n)$ bits), and $t$ represents the contents of the work tape $(s(n)$ bits). The configuration graph $G$ of $T$ is a directed acyclic graph with $2^{g(n)}$ nodes, one for each configuration of $T$, and an edge from $u$ to $v$ if $T$ can move in one step from configuration $u$ to configuration $v$. We include self-loops in this graph.

As in the full version of \cite{muggles}, let $B_x$ denote the adjacency matrix of $T$. The circuit $C$ first computes the entries of $B_x$, and then computes $\log t(n)$ matrices $B_{\log t(n)}, \dots, B_0$, where the $(u, v)$'th entry of $B_p$ is 1 if there is a path of length at most $2^{\log t(n) - p}$ from $u$ to $v$ in $G$. The matrix $B_p$ is obtained from $B_{p+1}$ by a process resembling repeated squaring of $B_x$ using naive matrix multiplication.\footnote{More specifically, $B_{p}[u, v] = 1 + \prod_{w \in \{0,1\}^{g(n)}} \left(1 + B_p[u, w] B_p[w, v]\right),$ where all arithmetic is done over an extension field of $\mathbb{F}_2$.} The wiring structure of this stage of the circuit is similar to that for naive matrix-vector multiplication, and it is straightforward to observe that the multilinear extensions of $add_i$ and $mult_i$ for these layers can be evaluated in $O(\log S(n))$ time and using $O(\log S(n))$ words of space. We omit these details for brevity.

\paragraph{Multilinear Extension of the Remaining Layer}
Thus, we need only show that the multilinear extensions of the wiring predicate of the layer of $C$ computing the entries of $B_x$ can be evaluated using $O(\log S(n))$ words of memory and $O(n\cdot s(n) \log n)$ time. Assume that $C$ has a designated input gate whose value is set to 0, and another whose value is set to 1; we call these the constant-0 and constant-1 input gates, respectively.
In determining the value of $B_x[u, v]$, the full version of \cite{muggles} demonstrates that there are 4 cases to consider. Notice configuration $u$ only reads one input bit, bit $x_i$.
\begin{enumerate}
\item Configuration $u$ can always go to $v$, regardless of $x_i$. Then $B_x[u, v]=1$.
\item Configuration $u$ can never to go $v$, regardless of $x_i$. Then $B_x[u, v]=0$.
\item Configuration $u$ can go to $v$ only if $x_i=1$. Then $B_x[u, v]=x_i$.
\item Configuration $u$ can go to $v$ only if \graham{$x_i=0$}. Then $B_x[u, v]=1+x_i$, with arithmetic done over an extension field of $\mathbb{F}_2$.
\end{enumerate}

Thus, all gates are layer $d(n)-1$ of $C$ are addition gates. In Case 1 above, the first input to gate $(u, v)$ is the constant-0 input gate, while the second is the constant-1 input gate. In Case 2, both inputs to gate $(u, v)$ at layer $d(n)-1$ equal the constant-0 input gate. In Case 3, the first input to gate ($u, v)$ is the constant-0 input gate, and the second input to gate $(u, v)$ is the $i$'th input gate. In Case 4, the first input to gate $(u, v)$ is the constant-1 input gate, and the second input to gate $(u, v)$ is the $i$'th input gate.

\balance
Write $u=(q_1, i_1, j_1, t_1)$ and $v=(q_2, i_2, j_2, t_2)$. The fact that the multilinear extension $\tilde{\text{add}}_{d(n)}(p, \omega_1, \omega_2)$ can be evaluated using $O(\log S(n))$ words of memory and time $O(n \cdot s(n) \log n)$ relies on the fact that \emph{computation is local}. More specifically, determining which of the four cases $(u, v)$ is in depends only on the states $q_1$ and $q_2$ (of which there are only $O(1)$ possibilities), the value of the $j_1$th bit in both $t_1$ and $t_2$ (of which there are only 4 possibilities),
determining whether the work-tape head can move $j_2-j_1$ locations to the right (this requires $j_2-j_1 \in \{-1, 0, 1\}$, and hence there are only $O(s(n))$ valid possibilities for $j_2$ and $j_1$), determining whether the input-tape head can move $i_2-i_1$ locations to the right (this requires $i_2-i_1 \in \{-1, 0, 1\}$, and hence there are only $O(n)$ valid possibilities for $i_2$ and $i_1$), and determining whether all other entries of $t_1$ and $t_2$ are identical (the multilinear extension of this predicate is succinct).

For example, $p$ is in Case 1 iff 
\begin{enumerate}
\item All bits of $t_1$ and $t_2$ other than bit $j_1$ are equal, and 
\item Given state $q_1$ and the value read by the work-tape head $t_{1, j_1}$, it holds that no matter the value of $x_i$, the non-deterministic machine can move to state $q_2$, move its output-tape head $j_2-j_1$ positions to the right, set bit $j_1$ of its work tape equal to $t_{2, j_1}$, and move its input head move $i_2-i_1$ positions to the right.
\end{enumerate} Let $S$ be the set of all values 
$(q_1, q_2, i_1, i_2, j_1, j_2, t_{1, j_1}, t_{2, j_1}) \in \{0, 1\}^{O(s(n))}$ values satisfying Property Two above. Notice all elements of $S$ can be enumerated in time $O(n \cdot s(n))$.

For $p=(q_1, i_1, j_1, t_1, q_2, i_2, j_2, t_2) \in \{0, 1\}^{2g(n)}$, $\omega_1, \omega_2 \in \{0, 1\}^{\log n}$, consider the multilinear polynomial 

$$\chi_{\text{Case 1}}(p, \omega_1, \omega_2) = \left(\sum_{x \in S} \chi_x(p_S)\right) \chi_{\bar{S}}(p_{\bar{S}}) \jedit{\chi_0(\omega_1) \chi_1(\omega_2)},$$ 

\noindent where $p_S$ denotes the vector $p$ restricted the entries corresponding to elements in $S$, $\chi_x$ is the multilinear polynomial testing for equality with $x$, $\chi_{\bar{S}}$ is the multilinear polynomial for testing that all bits of $t_1$ and $t_2$ other than $i_1$ are equal, \jedit{$\chi_0$ is the multilinear polynomial for testing that $\omega_1$ is equal to the index of the constant-0 gate, and $\chi_1$ is the multilinear polynomial for testing that $\omega_2$ is equal to the index of the constant-1 gate.}
This polynomial can clearly be evaluated in in time $O(n \cdot s(n) \log n)$ using $O(\log S(n))$ words of memory, and it evaluates to 1 on boolean input $(p, \omega_1, \omega_2)$ to 1 if $p$ is in Case 1 and $\omega_1$ and $\omega_2$ are as required by Case 1, and evaluates to zero otherwise. 

Similar polynomials $\chi_{\text{Case 2}}, \chi_{\text{Case 3}}, \chi_{\text{Case 4}}$ can be constructed for Cases 2-4. Thus, we can write 
$$\tilde{\text{add}_i}(p, \omega_1, \omega_2) = \sum_{j=1}^4 \chi_{\text{Case } j}(p, \omega_1, \omega_2),$$

which can clearly be evaluated in time $O(n \cdot s(n) \log n)$ and space $O(\log S(n))$.

\eat{
\section{Discussion}

As mentioned above, we believe that the restriction to arithmetic circuits which are amenable to the use of multilinear extensions
is not overly restrictive in practice. Indeed, all of the circuits we experimented on are amenable in this way, and any circuit that
isn't amenable would have to have an irregular wiring, which would
make such a circuit difficult to implement. We give a sufficient condition for a circuit to be amenable to the use of multilinear
extensions in Theorem \ref{thm:formal}.
It remains an interesting theoretical question whether there is a protocol
for all of NC in which the prover runs in $\tilde{O}(S(n))$ time, and the
verifier uses logarithmic space \emph{and} nearly linear time. 
 also the discussion section at the end of this note).
}

\clearpage
{

}

\end{document}